\documentclass[aps,showpacs,preprintnumbers,nofootinbibt,twocolumn]{revtex4}
\usepackage{eurosym}
\usepackage{amssymb}
\usepackage{graphicx}
\usepackage{color}

\def\be{\begin{equation}}
\def\ee{\end{equation}}
\def\bea{\begin{eqnarray}}
\def\eea{\end{eqnarray}}

\begin{document}

\title{Superconducting dark energy}
\author{Shi-Dong Liang$^1$}
\email{stlsd@sysu.edu.cn}
\author{Tiberiu Harko$^{2}$}
\email{t.harko@ucl.ac.uk}
\affiliation{$^1$State Key Laboratory of Optoelectronic Material and Technology, and
Guangdong Province Key Laboratory of Display Material and Technology, School
of Physics and Engineering,\\
Sun Yat-Sen University, Guangzhou 510275, People's Republic of China}
\affiliation{$^2$ Department of Mathematics, University College London, Gower Street,
London, WC1E 6BT, United Kingdom}
\date{\today }

\begin{abstract}
Based on the analogy with superconductor physics we consider a
scalar-vector-tensor gravitational model, in which the dark energy action is
described by a gauge invariant electromagnetic type functional. By assuming
that the ground state of the dark energy is in a form of a condensate with
the U(1) symmetry spontaneously broken, the gauge invariant electromagnetic
dark energy can be described in terms of the combination of a vector and of
a scalar field (corresponding to the Goldstone boson), respectively. The
gravitational field equations are obtained by also assuming the possibility
of a non-minimal coupling between the cosmological mass current and the
superconducting dark energy. The cosmological implications of the dark
energy model are investigated for a Friedmann-Robertson-Walker homogeneous
and isotropic geometry for two particular choices of the electromagnetic
type potential, corresponding to a pure electric type field, and to a pure
magnetic field, respectively. The time evolutions of the scale factor,
matter energy density and deceleration parameter are obtained for both
cases, and it is shown that in the presence of the superconducting dark
energy the Universe ends its evolution in an exponentially accelerating
vacuum de Sitter state. By using the formalism of the irreversible
thermodynamic processes for open systems we interpret the generalized
conservation equations in the superconducting dark energy model as
describing matter creation. The particle production rates, the creation
pressure and the entropy evolution are explicitly obtained.
\end{abstract}

\maketitle

%




\section{Introduction}

\label{Sect.I}

The $\Lambda $ Cold Dark Matter ($\Lambda $CDM) model of cosmology is
remarkably successful in accounting for most of the observed features of the
Universe. The recent Planck satellite data from the 2.7 degree Cosmic
Microwave Background full sky survey \cite{P1,P2} have generally confirmed
again the present day standard cosmological paradigm. However, a number of fundamental
questions at the very foundations of cosmology and gravitation still remain
open, and unanswered. Perhaps the most important challenge facing modern cosmology is the
understanding of the mechanism of the acceleration of the late universe, which is usually attributed to the presence of the mysterious dark energy. \cite{latetime}. In
fact, as fundamental candidates responsible for the cosmic expansion, the
standard $\Lambda $CDM model of cosmology has favored dark energy models
involving time-dependent scalar fields. Scalar fields naturally arise in many
particle physics models, including string theory. On the other hand, the
underlying dynamics of inflationary models, assumed to be of fundamental importance for
the understanding of the early history of the Universe, also depend essentially on a
single scalar field, the inflaton, rolling in some underlying potential \cite%
{1}. The possibility that a single canonical scalar field $\phi$, with a non-zero
potential, called \textit{quintessence}, could be responsible for the
late-time cosmic acceleration, was also explored in much detail \cite{Q1,Q2,caldwell}. The well-known 
action for a scalar field in the presence of gravity is
\begin{equation}  \label{acts}
S_{\phi}=\int{\left[\frac{R}{16\pi G} -\frac{1}{2}\nabla ^{\alpha }\phi
\nabla _{\alpha }\phi -V(\phi )\right]\sqrt{-g}d^4x},
\end{equation}
where $R$ is the Ricci scalar, $G$ is the gravitational constant, and $%
V(\phi )$ is the self-interaction potential, respectively \cite{Fa04}.

In opposition to the behavior of the cosmological constant, the quintessence
equation of state changes dynamically with time \cite{11}. In fact,
many other exotic fluids have been proposed to explain the accelerated
expansion of the Universe. Some of the proposed models are the so-called $k-$essence
models, in which the late-time acceleration is driven by the kinetic energy term of the
scalar field \cite{kessence}. A number of coupled models,  where dark energy interacts both quantitatively and qualitatively 
with dark matter, have also been proposed \cite{coupledDE}, as well as unified models of dark matter and
dark energy \cite{DM_DE}. For a review of the dark energy candidates see \cite{Copeland:2006wr}.

An intriguing alternative about the nature of dark energy, which was also
intensively investigated in the literature, is the possibility that it could
be described by a vector field, which can be at the origin of the present
stage of cosmic acceleration. In its simplest formulation the action for
the vector field dark energy  model is
\begin{eqnarray}  \label{actv}
S_{V}&=&\int d^4x \sqrt{-g}\Bigg\{\frac{R}{16\pi G} -\sum_{a=1}^{3}\left[%
\frac{1}{4}F^a_{\mu\nu}F^a{}^{\mu\nu} +V(A^2)\right]+  \nonumber \\
&&L_m\Bigg\},
\end{eqnarray}
where $F_{\mu\nu}^a=\partial_\mu A^a_\nu-\partial_\nu A^a_\mu$, $%
A^2=g^{\mu\nu}A^a_\mu A^a_\nu$, and $L_m$ is the matter Lagrangian \cite{v1}%
. This vector (or more exactly Yang-Mills) type action for the dark energy thus contains three identical
components obtained by generalizing the Lagrangian of a single vector field. The term $V(A{}^2)$ is a
self-interaction potential that explicitly violates gauge invariance. The cosmological
implications of the vector type dark energy models have been investigated in
\cite{v2}.

More general vector field dark energy models, in which the vector field is
non-minimally coupled to the gravitational field, have been proposed in \cite%
{v3}. By assuming that the Universe is filled with a massive cosmological
vector field, with mass $\mu _{\Lambda }$, which is characterized by a
four-potential $\Lambda ^{\mu }\left( x^{\nu }\right) $, $\mu,\nu =0,1,2,3$,
which couples non-minimally to gravity, and by introducing, in analogy with
electrodynamics, the field tensor $C_{\mu \nu }=\nabla _{\mu } \Lambda _{\nu
}-\nabla _{\nu } \Lambda_{\mu }$, the action for the non-minimally coupled
vector dark energy theory can be written as
\begin{eqnarray}  \label{1}
S&=&-\int \Bigg[ R+C_{\mu \nu }C^{\mu \nu }+\frac{1}{2}\mu _{\Lambda
}^{2}\Lambda _{\mu }\Lambda ^{\mu }+\omega \Lambda _{\mu }\Lambda ^{\mu }R+
\nonumber \\
&&\eta \Lambda ^{\mu }\Lambda ^{\nu }R_{\mu \nu }+16\pi G_0L_{m}\Bigg] \sqrt{%
-g}d\Omega ,
\end{eqnarray}
where $R_{\mu \nu }$ is the Ricci tensor and $G_0$ is the gravitational
constant. In Eq.~(\ref{1}) $\omega $ and $\eta $ are dimensionless coupling
parameters.

At first sight the gravitational actions given by Eqs.~(\ref{acts}) and (\ref%
{actv}) look totally different, from both mathematical point of view, as
well as from the physical interpretation point of view. However, they can be
in fact interpreted and understood as the limiting cases of a single
physical model, related to the spontaneous breaking of the electromagnetic
U(1) symmetry. Thus an approach is used to describe superconductivity from a
fundamental point of view \cite{W,Cas,F}.

From a general physical point of
view in the theory of superconductivity the existence of a quantum
condensate (superconducting state) is described by a non-vanishing value of
a gauge dependent complex order parameter \cite{W,Cas,F}. In bosonic systems superfluid
behavior occurs when the expectation value of the bosonic field parameter $%
\psi $ has a nonzero value, $\left \langle \psi \right
\rangle \neq 0$. On
the other hand the existence of superconductivity is also induced by a nonzero
value of the expectation value of the pair field operator. Therefore, in the
ground state of a superconducting system a quantum condensate $%
\left
\langle \epsilon _{\alpha \beta }\psi ^{\alpha }\psi ^{\beta
}\right
\rangle $ forms \cite{W,Cas,F}. Since the difermion operator has
charge $-2e$, the important result that the quantum condensate breaks the electromagnetic U(1)
symmetry is obtained. Another fundamental quantity  in the model is a scalar field, $\Phi $,
which plays the role of the order parameter. Under a gauge
transformation $A_{\mu}\rightarrow A_{\mu}+\partial _{\mu}\Lambda $, the scalar field
transforms like the condensate wave function $\psi \rightarrow e^{ie\Lambda }\psi
\Longrightarrow \Phi \rightarrow e^{2ie\Lambda }\Phi$. Note that in the
zero temperature superconductivity theory one also introduces the Goldstone field $\phi $ as the phase of the field $\Phi$, $\Phi=\rho e^{2ie\phi}$, as
well as the gauge invariant Fermi fields, $\tilde{\psi}=e^{-ie\phi} $ \cite%
{W,Cas}.

Hence from a fundamental point of view a superconducting system can be described by a gauge invariant
Lagrangian, depending on the wave function $\tilde{\psi} $, and on the vector potentials $A_{\mu}$ and $\nabla _{\mu }\phi$.
A simplified model is obtained after integrating out the Fermi fields. Thus one obtains a gauge invariant Lagrangian,
depending only on $A_{\mu}$ and $\nabla _{\mu}\phi$, respectively. The important requirement of the gauge
invariance of the theory implies that these bosonic fields must appear only
in the combinations $F_{\mu \nu}=\nabla _{\mu}A_{\nu}-\nabla _{\nu}A_{\mu}$
and $A_{\mu}-\nabla _{\mu}\phi$, respectively. Therefore the Lagrangian
describing a superconductor from a fundamental physical point of view has
the form \cite{W,Cas}
\begin{equation}  \label{l1}
L=-\frac{1}{4}\int{F_{\mu \nu}F^{\mu \nu}d^3\vec{r}}+L_s\left(A_{\mu}-\nabla
_{\mu}\phi\right),
\end{equation}
where $L_s\left(A_{\mu}-\nabla _{\mu}\phi\right)$ is an arbitrary function
of the argument $A_{\mu}-\nabla _{\mu}\phi$. The only physical condition
required on the superconductor Lagrangian $L_s$ is that in the absence of $%
A^{\mu}$ and $\phi $ it gives rise to a stable state of the system. In
particular, this requires that the point $A_{\mu} = \nabla _{\mu}\phi$ is a
local minimum of the theory (this property can fully explain the Meissner effect in
superconductivity theory \cite{W}). Therefore, we require that the second derivative of the
superconductor Lagrangian $L_s$ with respect to its argument must be nonzero
at the point $A_{\mu} = \nabla _{\mu}\phi$  \cite{W,Cas}.

It is the goal of the present paper to consider a gravitational model in
which dark energy is described by a Lagrangian of the form given by Eq.~(\ref%
{l1}), resulting from the breaking of the U(1) symmetry in the ground state
dark energy condensate. By analogy with condensed matter physics we call
this model the \textit{superconducting dark energy} model. The gravitational
field equations of the model are derived from an action principle, and the
cosmological implications are investigated in a background homogeneous and
isotropic flat Friedmann-Robertson-Walker geometry. We consider two distinct
classes of cosmological models, corresponding to two different choices of
the electromagnetic potential $A_{\mu}$ of the dark energy. In the first
model $A_{\mu}$ has only a non-vanishing temporal component, while in the
second case we assume non-vanishing spatial (magnetic) components of the
potential. In both cases we assume that the dark energy self-interaction
potential is constant.

In the present dark energy model, due to the coupling between the matter
current and the electromagnetic and scalar potentials of the dark energy,
the matter energy-momentum tensor is not conserved. By using the formalism
of the open thermodynamic systems introduced in \cite{Pri0}-\cite{Lima} (see
also \cite{ref} for recent investigations of particle creation in
cosmology), we interpret the generalized conservation equations in the
superconducting dark energy model from a thermodynamic point of view as
describing irreversible matter creation processes. Thus in the present model
particle creation corresponds to an irreversible energy flow from the
superconducting dark energy to the created matter constituents (both normal
and dark). We explicitly obtain the equivalent particle number creation
rates, the creation pressure and the entropy production rates. The
temperature evolution laws of the newly created particles are explicitly
derived. We also show that due to the superconducting dark energy - matter
current coupling, during the cosmological evolution a large amount of
comoving entropy could be produced.

The present paper is organized as follows. In Section~\ref{sect1} the
gravitational field equations of the superconducting dark energy model, a
scalar-vector-tensor theory with broken U(1) symmetry, are derived from a
variational principle. The equations of motion of the scalar and vector
fields are also obtained. The cosmological applications of the theory are
investigated in Section~\ref{sect2}. Two distinct dark energy models are
considered: an electric type, in which the vector potential has only a time
component, and a magnetic type, with the vector potential having only
spatial components. The cosmological properties of both models are
investigated in detail. The thermodynamic interpretation of the
superconducting dark energy model is considered, in the framework of the
thermodynamic of open systems and irreversible processes, in Section \ref%
{therm}. We discuss and conclude our results in Section~\ref{sect3}. In this
paper we adopt the Landau-Lifshitz \cite{LaLi} metric conventions, and we
use the natural system of units with $8\pi G=c=1$.

\section{Field equations of the superconducting dark energy model}

\label{sect1}

In the following we assume that the interaction of the gravitational and of the
superconducting dark energy scalar-vector fields is described by a
Lagrangian which is required to satisfy the following standard conditions: a) the
Lagrangian density is a four-scalar b) the free-field energies are
positive-definite for all the metric, scalar and vector fields c) the
resulting theory is metric and d) the field equations contain no higher than
second order derivatives of the fields~\cite{hel73}. Based on the analogy
with superconductor physics we consider a gravitational
scalar-vector-tensor action of the form
\begin{eqnarray}  \label{s1}
S&=&-\int \Bigg[ \frac{R}{2}+\frac{1}{16\pi }F_{\mu \nu }F^{\mu \nu }-\frac{%
\lambda}{2} g^{\mu \nu }\times  \nonumber \\
&&\left( A_{\mu }-\nabla _{\mu }\phi \right) \left( A_{\nu }-\nabla _{\nu
}\phi \right) +V\left( A^2,\phi \right) -  \nonumber \\
&&\frac{\alpha}{2} g^{\mu \nu }j_{\mu }\left( A_{\nu }-\nabla _{\nu }\phi
\right) +L_{m}\left( g_{\mu \nu },\psi \right) \Bigg] \sqrt{-g}d\Omega ,
\end{eqnarray}%
where $\lambda $ and $\alpha $ are constants, $L_{m}\left( g_{\mu \nu },\psi
\right) $ is the Lagrangian of the total (ordinary baryonic plus dark)
matter, and $j^{\mu }=\rho u^{\mu }$ is the total mass current, where $\rho $
is the total matter density (including dark matter), and $u^{\mu }$ is the
matter four-velocity. We assume that the baryonic and dark matter are
comoving. The third term in the action Eq.~(\ref{s1}) follows from the
assumption that the superconducting dark energy is close to the minimum $%
A_{\mu}=\nabla _{\mu}\phi$. In this case the general superconducting
Lagrangian (\ref{l1}) can be expanded in power series as \cite{W,Cas}
\begin{equation}
L_s\left(A_{\mu}-\nabla _{\mu}\phi \right)\approx L_0+\frac{1}{2}\frac{%
\delta ^2L_s}{\delta \left(A_{\mu}-\nabla _{\mu}\phi\right)^2}%
\left(A_{\mu}-\nabla _{\mu}\phi\right)^2+...,
\end{equation}
where $L_0$ is a constant. Hence the superconducting type Lagrangian $%
L_s\left(A_{\mu}-\nabla _{\mu}\phi \right)$ gives a quadratic contribution
in $A_{\mu}-\nabla _{\mu}\phi$ to the gravitational Lagrangian. We have also
assumed the possibility of an interaction between the total matter flux $%
j^{\mu}$ and the superconducting dark energy gauge invariant potentials $%
A_{\mu}-\nabla _{\mu}\phi$. $V(A^2,\phi)$ is the self-interaction potential
of the scalar and vector fields, with $A^2=A_{\mu}A^{\mu}$, in which we have
also included the constant $L_0$. When $\phi \equiv 0$, that is, the scalar
field vanishes, the action (\ref{s1}) gives the pure vector model of the
dark energy. When the electromagnetic type potential $A^{\mu}=0$, we recover
the standard action of the minimally coupled scalar-tensor theory. Hence the
gravitational action (\ref{s1}) gives a unified framework for the minimal
inclusion into the gravitational action of the scalar-vector interactions,
under the assumption of the existence of a U(1) broken symmetry. The second
and third terms in the gravitational action are also similar to the
Stueckelberg Lagrangian \cite{Stue}.

We define the energy-momentum tensor of the matter as \cite{LaLi}
\begin{equation}
T_{\mu \nu }=-\frac{2}{\sqrt{-g}}\left[ \frac{\partial \left( \sqrt{-g}%
L_{m}\right) }{\partial g^{\mu \nu }}-\frac{\partial }{\partial x^{\lambda }}%
\frac{\partial \left( \sqrt{-g}L_{m}\right) }{\partial \left( \partial
g^{\mu \nu }/\partial x^{\lambda }\right) }\right] .
\end{equation}

By making the important assumption that the Lagrangian density $L_{m}$ of the matter depends only
on the metric tensor components $g_{\mu \nu }$, and not on its derivatives,
we obtain the expression $T_{\mu \nu }=L_{m}g_{\mu \nu }-2\partial L_{m}/\partial g^{\mu
\nu }$.

By varying the action (\ref{s1}) with respect to the metric tensor we obtain
the gravitational field equations for the superconducting dark energy model
as
\begin{eqnarray}  \label{feq}
&&\hspace*{-5mm}R_{\mu \nu}-\frac{1}{2}Rg_{\mu \nu}=T_{\mu \nu}+ \frac{1}{%
4\pi}\left(-F_{\mu \alpha }F^{\alpha }_{\nu}+\frac{1}{4}F_{\alpha \beta
}F^{\alpha \beta }g_{\mu \nu}\right)+  \nonumber \\
&&\hspace*{-5mm}\lambda \left(A_{\mu}-\nabla _{\mu}\phi
\right)\left(A_{\nu}-\nabla _{\nu}\phi \right)- \frac{\lambda }{2}%
\left(A^{\alpha }-\nabla ^{\alpha }\phi \right)\left(A_{\alpha }-\nabla
_{\alpha }\phi \right)\times  \nonumber \\
&&\hspace*{-5mm}g_{\mu \nu}+ \alpha j_{\mu}\left(A_{\nu}-\nabla _{\nu}\phi
\right)-\frac{\alpha }{2}j^{\beta }\left(A_{\beta }-\nabla _{\beta
}\phi\right)g_{\mu \nu}+  \nonumber \\
&&\hspace*{-5mm}V\left(A^2,\phi\right)g_{\mu \nu},
\end{eqnarray}
where $T_{\mu \nu}$, the energy-momentum tensor of the ordinary matter, is
given by
\begin{equation}  \label{9}
T_{\mu \nu}=\left(\rho +p\right)u_{\mu}u_{\nu}-pg_{\mu \nu},
\end{equation}
where $p$ is the total thermodynamic pressure of the matter components
(baryonic and dark). By taking the variation of the action Eq.~(\ref{s1})
with respect to the scalar field $\phi $ we obtain
\begin{eqnarray}  \label{var1}
\delta _{\phi}S&=&- \int\Bigg[\lambda A^{\mu}\nabla _{\mu}\delta \phi
-\lambda g^{\mu \nu}\nabla _{\mu}\delta \phi \nabla _{\nu}\phi +  \nonumber
\\
&&\frac{\alpha }{2}j^{\mu}\nabla _{\mu}\delta \phi +\partial
_{\phi}V\left(A^2,\phi\right)\delta \phi \Bigg]\sqrt{-g} d\Omega.
\end{eqnarray}
With the use of the mathematical identity
\begin{equation}
\nabla _{\mu}\left(B^{\mu}\delta \phi\right)=\nabla _{\mu}B^{\mu}\delta \phi
+B^{\mu}\nabla _{\mu}\delta \phi,
\end{equation}
it follows that if the conditions
\begin{equation}  \label{cond1}
\nabla _{\mu}A^{\mu}=0,\nabla _{\mu}j^{\mu}=0,
\end{equation}
are imposed on the dark energy vector potential, and on the baryonic matter
flow, the variation in the integral (\ref{var1}) of the first term,
containing $A^{\mu}$, and of the last term, containing $j^{\mu}$, vanish
identically. Therefore we obtain the result that if the conditions given by
Eqs.~(\ref{cond1}) are satisfied, then the scalar field satisfies the
standard Klein-Gordon equation,
\begin{equation}
\lambda g^{\mu \nu}\nabla _{\mu}\nabla _{\nu}\phi +\partial
_{\phi}V\left(A^2,\phi\right)=0.
\end{equation}
This case corresponds to the minimal coupling of the scalar and vector
fields. However, in the following we will use a more general approach, in
which no additional constraints are imposed on the fields or on the
hydrodynamic flow. Therefore, the variation of the action Eq.~(\ref{s1})
gives the following coupled evolution equation for the scalar and vector
fields,
\begin{equation}
\lambda g^{\mu \nu}\nabla _{\mu}\nabla _{\nu}\phi +\partial
_{\phi}V\left(A^2,\phi\right)-\lambda \nabla _{\mu}A^{\mu}-\frac{\alpha }{2}%
\nabla _{\mu}j^{\mu}=0.
\end{equation}

By varying the superconducting dark energy action Eq.~(\ref{s1}) with
respect to $A_{\mu}$, we obtain first
\begin{eqnarray}
\delta _{A^{\mu }}S&=&\int \Bigg[ -\frac{1}{4\pi }F^{\mu \nu }\nabla _{\nu
}\delta A_{\mu }-\lambda g^{\mu \nu }\left( A_{\mu }-\nabla _{\mu }\phi
\right) \delta A_{\mu }-  \nonumber \\
&&\frac{\alpha}{2} j^{\mu }\delta A_{\mu }+2\partial _{A^{2}}V\left(
A^{2},\phi \right) A^{\mu }\delta A_{\mu }\Bigg] \sqrt{-g}d\Omega =0.
\nonumber \\
\end{eqnarray}

By taking into account the identity $\nabla _{\nu }\left( F^{\mu \nu }\delta
A_{\mu }\right) =\nabla _{\nu }F^{\mu \nu }\delta A_{\mu }+F^{\mu \nu
}\nabla _{\nu }\delta A_{\mu }$, after partial integration and the use of
Gauss' theorem, it follows that the superconducting dark energy vector field
satisfies the equation
\begin{equation}
\frac{1}{4\pi }\nabla _{\nu }F^{\mu \nu }=J^{\mu},
\end{equation}
where
\begin{equation}
J^{\mu}=\left[\lambda g^{\mu \nu }\left( A_{\nu }-\nabla _{\nu }\phi \right)
+\frac{\alpha}{2} j^{\mu }-2\partial _{A^{2}}V\left( A^{2},\phi \right)
A^{\mu }\right].
\end{equation}
The divergence of the dark energy field tensor can be obtained as $\nabla
_{\nu }F^{\mu \nu }=\left(1/\sqrt{-g}\right)\partial _{\nu}\left(\sqrt{-g}%
F^{\mu\nu}\right)$.

By its definition the dark energy electromagnetic type tensor $F^{\mu \nu }$
satisfies the Bianchi identity
\begin{equation}
\varepsilon ^{\alpha \beta \mu \nu }\nabla _{\beta }F_{\mu \nu }=0,
\end{equation}
where $\varepsilon ^{\alpha \beta \mu \nu }$ is the complete antisymmetric
unit tensor of rank four.

Finally, by taking the covariant derivative of the field equations Eqs.~(\ref%
{feq}) we obtain the matter conservation equation in the presence of a
superconducting dark energy as
\begin{eqnarray}  \label{consem}
&&\nabla _{\mu }T_{\nu }^{\mu }+\frac{\alpha}{2} \nabla _{\mu }\left[j^{\mu
}\left( A_{\nu }-\nabla _{\nu }\phi \right)\right] -\frac{\alpha }{2}\nabla
_{\nu }j^{\beta }\left( A_{\beta }-\nabla _{\beta }\phi \right) +  \nonumber
\\
&&\partial _{\phi }V\left( A^{2},\phi \right) A_{\nu }+ 2\partial
_{A^{2}}V\left( A^{2},\phi \right) A^{\alpha }\nabla _{\alpha }A_{\nu }=0.
\end{eqnarray}
The derivation of Eq.~(\ref{consem}) is presented in Appendix \ref{app1}.

By taking into account the explicit form of the energy-momentum tensor,
given by Eq.~(\ref{9}) we obtain
\begin{eqnarray}  \label{cem1}
&&\hspace{-0.5cm}\left(\nabla ^{\mu }\rho +\nabla ^{\mu} p\right)u_{\mu
}u_{\nu }+\left(\rho +p\right)u_{\nu}\nabla ^{\mu} u_{\mu} + \left(\rho
+p\right)u_{\mu}\nabla ^{\mu }u_{\nu}-  \nonumber \\
&&\hspace{-0.5cm}\nabla ^{\mu }pg_{\mu \nu}+\frac{\alpha}{2} \nabla _{\mu }%
\left[j^{\mu }\left( A_{\nu }-\nabla _{\nu }\phi \right)\right] - \frac{%
\alpha }{2}\nabla _{\nu }j^{\beta }\left( A_{\beta }-\nabla _{\beta }\phi
\right) +  \nonumber \\
&&\hspace{-0.5cm}\partial _{\phi }V\left( A^{2},\phi \right) A_{\nu }+
2\partial _{A^{2}}V\left( A^{2},\phi \right) A^{\alpha }\nabla _{\alpha
}A_{\nu }=0.
\end{eqnarray}

By multiplying Eq.~(\ref{cem1}) with $u^{\nu }$, and by taking into account
the mathematical identity $u^{\nu }\nabla ^{\mu }u_{\nu }=0$ we obtain the
energy conservation equation in the superconducting dark energy model as
\begin{eqnarray}  \label{em2}
&&\dot{\rho}+3\left( \rho +p\right) H+\frac{\alpha }{2}u^{\nu }\nabla _{\mu }%
\left[ j^{\mu }\left( A_{\nu }-\nabla _{\nu }\phi \right) \right] -
\nonumber \\
&&\frac{\alpha }{2}\frac{d}{ds}\left[j^{\beta }\left( A_{\beta }-\nabla
_{\beta }\phi \right)\right] + \partial _{\phi }V\left( A^{2},\phi \right)
u^{\nu }A_{\nu }+  \nonumber \\
&&2\partial _{A^{2}}V\left( A^{2},\phi \right)u^{\nu } A^{\alpha }\nabla
_{\alpha }A_{\nu }=0,
\end{eqnarray}
where we have introduced the Hubble function $H=(1/3)\nabla ^{\mu }u_{\mu }$%
, and we have denoted $\dot{}=u^{\mu }\nabla _{\mu }=d/d{s}$, respectively,
where $d{s}$ is the line element corresponding to the metric $g_{\mu \nu }$,
$d{s}^{2}=g_{\mu \nu }dx^{\mu }dx^{\nu }$.

By multiplying Eq.~(\ref{cem1}) with the projection operator $h_{\lambda
}^{\nu }$, defined as $h_{\lambda }^{\nu }=\delta _{\lambda }^{\nu
}-u_{\lambda }u^{\nu }$, and satisfying the relation $u_{\nu }h_{\lambda
}^{\nu }=0$, gives the momentum balance equation for a perfect fluid in the
superconducting dark energy model as
\begin{eqnarray}  \label{force}
&&u^{\mu }\nabla _{\mu }u^{\lambda }=\frac{d^{2}x^{\lambda }}{ds^{2}}+\Gamma
_{\mu \nu }^{\lambda }u^{\mu }u^{\nu }=\frac{h^{\nu \lambda }}{\rho +p}%
\Bigg\{ \nabla _{\nu }p-  \nonumber \\
&&\frac{\alpha }{2}\nabla _{\mu }\left[ j^{\mu }\left( A_{\nu }-\nabla _{\nu
}\phi \right) \right] +\frac{\alpha }{2}\nabla _{\nu }j^{\beta }\left(
A_{\beta }-\nabla _{\beta }\phi \right) -  \nonumber \\
&&\partial _{\phi }V\left( A^{2},\phi \right) A_{\nu }-2\partial
_{A^{2}}V\left( A^{2},\phi \right) A^{\alpha }\nabla _{\alpha }A_{\nu }%
\Bigg\} .
\end{eqnarray}

\section{Cosmological applications}\label{sect2}

We assume that the metric of the Universe is given by the isotropic and
homogeneous Friedmann-Robertson-Walker metric,
\begin{equation}
ds^2=dt^2-a^2(t)\left(dx^2+dy^2+dz^2\right),
\end{equation}
where $a(t)$ is the scale factor describing the expansion of the Universe.
We assume that the cosmological matter is comoving with the cosmological
expansion, and therefore we choose the four velocity of the cosmological
fluid as $u^{\mu}=(1,0,0,0)$. Hence the components of the four-current
vector are $j^{\mu}=\left(\rho,0,0,0\right)$. In the
Friedmann-Robertson-Walker geometry the Hubble function takes the form $H=%
\dot{a}/a$, since $u^{\mu }\nabla _{\mu }=\dot{}=d/dt$. To describe the
decelerating/accelerating nature of the cosmological expansion, we use the
deceleration parameter $q$, with the definition
\begin{equation}
q=\frac{d}{dt}\frac{1}{H}-1=-\frac{\dot{H}}{H^2}-1.
\end{equation}

Moreover, from the homogeneity of the Universe it follows that the scalar
and vector fields $\phi $ and $A_{\mu}$ can be only functions of the
cosmological time $t$, so that $\phi =\phi (t)$ and $A_{\mu}=A_{\mu}(t)=%
\left(A_0(t),A_1(t),A_2(t),A_3(t)\right)$, respectively. The non-zero
components of the dark energy tensor $F_{\mu \nu}$ are given by $F_{i0}(t)=-%
\dot{A}_i(t)$, $i=1,2,3$, and $F^{i0}(t)=\dot{A}_i(t)/a^2(t)$, $i=1,2,3$.
Hence we obtain $F_{\alpha \beta }F^{\alpha \beta
}=-\left(2/a^2(t)\right)\sum _{i=1}^3{\left[\dot{A}_i(t)\right]^2}$. Then,
the cosmological equations corresponding to the superconducting dark energy
model are given by
\begin{eqnarray}  \label{f1}
&&3H^2=\left[1+\frac{\alpha }{2}\left(A_0-\dot{\phi}\right)\right]\rho +
\frac{1}{8\pi a^2(t)}\left(\sum _{k=1}^3\dot{A}_k^2\right)+  \nonumber \\
&&\frac{\lambda }{2}\left(A_0-\dot{\phi}\right)^2+\frac{\lambda}{2a^2(t)}%
\left(\sum _{k=1}^3 {A_k^2}\right)+V\left(A^2,\phi\right),
\end{eqnarray}
\begin{eqnarray}  \label{f2}
&&\hspace*{-2mm}2\dot{H}+3H^2=-p+\frac{1}{4\pi}\frac{\dot{A}_i^2}{a^2(t)}-%
\frac{1}{8\pi a^2(t)}\left(\sum _{k=1}^3\dot{A}_k^2\right)-  \nonumber \\
&&\hspace*{-2mm}\lambda \frac{A_i^2}{a^2(t)}- \frac{\lambda }{2}\left(A_0-%
\dot{\phi}\right)^2+\frac{\lambda }{2a^2(t)}\left(\sum_{k=1}^3{A_k^2}\right)-
\nonumber \\
&&\frac{\alpha }{2}\left(A_0-\dot{\phi}\right)\rho + V\left(A^2,\phi\right),
i=1,2,3,
\end{eqnarray}
\begin{eqnarray}  \label{f3}
&&\hspace*{-13mm}\lambda \ddot{\phi}-3\left[\frac{\alpha }{2}\rho +\lambda
\left(A_0-\dot{\phi}\right)\right]H-\lambda \dot{A}_0-\frac{\alpha }{2}\dot{%
\rho}+  \nonumber \\
&&\hspace*{-13mm}\partial _{\phi}V\left(A^2,\phi\right)=0,
\end{eqnarray}
\begin{equation}
\hspace*{-13mm}\lambda \left(A_0-\dot{\phi}\right)+\frac{\alpha}{2}\rho
-2\partial _{A^2}V\left(A^2,\phi\right)A_0=0,
\end{equation}
\begin{eqnarray}
&&\ddot{A}_{k}+H\dot{A}_{k}(t)+4\pi \lambda A_{k}(t)-8\pi \partial
_{A^{2}}V\left( A^{2},\phi \right) A_{k}(t)=0,  \nonumber \\
&&k=1,2,3.
\end{eqnarray}

As an independent variable we introduce, instead of the cosmological time $t$%
, the redshift $z$, defined as $1+z=1/a$. Therefore
\begin{equation}
\frac{dH}{dt}=\frac{dH}{dz}\frac{dz}{dt}=-(1+z)H\frac{dH}{dz}.
\end{equation}
As a function of the redshift the deceleration parameter is obtained as
\begin{equation}
q=(1+z)\frac{1}{H(z)}\frac{dH(z)}{dz}-1.
\end{equation}

In the following we will explicitly investigate two distinct superconducting
dark energy models.

\subsection{Electric dark energy models}

We assume that the dark energy vector potential has the form $A_{\mu
}=\left( A_{0}(t),0,0,0\right) $, that is, the dark energy vector potential
has only one, electric type, component. For this choice $F_{\mu \nu }\equiv
0 $, $\forall \mu ,\nu \in \lbrack 0,1,2,3]$. The gravitational field
equations describing the cosmological dynamics in the presence of the
superconducting dark energy take the form
\begin{equation}
3H^{2}=\left[ 1+\frac{\alpha }{2}\left( A_{0}-\dot{\phi}\right) \right] \rho
+\frac{\lambda }{2}\left( A_{0}-\dot{\phi}\right) ^{2}+V\left( A^{2},\phi
\right) ,  \label{e1}
\end{equation}
\begin{equation}
2\dot{H}+3H^{2}=-p-\frac{\lambda }{2}\left( A_{0}-\dot{\phi}\right) ^{2}-%
\frac{\alpha }{2}\left( A_{0}-\dot{\phi}\right) \rho +V\left( A^{2},\phi
\right) ,  \label{e2}
\end{equation}
\begin{equation}
\frac{d}{dt}\left\{ a^{3}\left[ \lambda \left( A_{0}-\dot{\phi}\right) +%
\frac{\alpha }{2}\rho \right] \right\} -a^{3}\partial _{\phi }V\left(
A^{2},\phi \right) =0,  \label{e3}
\end{equation}
\begin{equation}
\lambda \left( A_{0}-\dot{\phi}\right) +\frac{\alpha }{2}\rho -2\partial
_{A^{2}}V\left( A^{2},\phi \right) A_{0}=0.  \label{e4}
\end{equation}

In order to close the system of equations (\ref{e1})-(\ref{e4}) the baryonic
equation of state $p=p\left( \rho \right) $ must also be provided. In the
following we will restrict our analysis to the case of a constant
self-interaction potential of the superconducting dark energy field, $%
V\left( A^{2},\phi \right) =V_0=\mathrm{constant}$. Then from Eqs.~(\ref{e3}%
) and (\ref{e4}) we obtain
\begin{equation}
\left( A_{0}-\dot{\phi}\right) =-\frac{\alpha }{2\lambda }\rho .
\end{equation}

Hence the generalized Friedmann equations of the cosmological expansion in
the presence of the electric type superconducting dark energy become
\begin{equation}
3H^{2}=\rho -\frac{\alpha ^{2}}{8\lambda }\rho ^{2} +V_0=\rho +\rho _{DE},
\label{hd0}
\end{equation}
\begin{equation}  \label{pd0}
2\dot{H}+3H^{2}=-p+\frac{\alpha ^{2}}{8\lambda }\rho ^{2}+V_0=-p-p_{DE},
\end{equation}%
where we have denoted
\begin{equation}
\rho _{DE}=-\frac{\alpha ^{2}}{8\lambda }\rho ^{2} +V_0,
\end{equation}
and
\begin{equation}
p_{DE}=-\frac{\alpha ^{2}}{8\lambda }\rho ^{2}-V_0,
\end{equation}
respectively. From Eqs.~(\ref{hd0}) and (\ref{pd0}) we obtain
\begin{equation}
2\dot{H}=-\left( \rho +p\right) +\frac{\alpha ^{2}}{4\lambda }\rho ^{2}.
\label{hd1}
\end{equation}

The energy conservation equation can be written as
\begin{equation}  \label{cex}
\frac{d}{dt}\left[ \left( \rho -\frac{\alpha ^{2}}{8\lambda }\rho
^{2}\right) a^{3}\right] +\left( p-\frac{\alpha ^{2}}{8\lambda }\rho
^{2}\right) \frac{d}{dt}a^{3}=0.
\end{equation}

The deceleration parameter can be obtained as
\begin{equation}
q=\frac{\left( \rho +3p\right)-\left( \alpha ^{2}/2\lambda \right) \rho
^{2}-2V_0}{2\left[\rho -\left( \alpha ^{2}/8\lambda \right) \rho ^{2}+V_0%
\right]}.
\end{equation}

For a dust Universe with $p=0$, the deceleration parameter takes the form
\begin{equation}
q=\frac{1}{2}\frac{\rho-\left( \alpha ^{2}/2\lambda \right) \rho^2-2V_0 }{%
\rho -\left( \alpha ^{2}/8\lambda \right) \rho ^2+V_0 }.
\end{equation}

From the field equations Eqs. (\ref{hd0})-(\ref{hd1}) we obtain the time
evolution equation of the baryonic matter density as
\begin{equation}  \label{rho1}
\dot{\rho}=-\sqrt{3}\frac{\sqrt{\rho -\left( \alpha ^{2}/8\lambda \right)
\rho ^{2}+V_{0}}\left\{ p+\rho \left[ 1-\left( \alpha ^{2}/4\lambda \right)
\rho \right] \right\} }{\left[ 1-\left( \alpha ^{2}/4\lambda \right) \rho %
\right] }.
\end{equation}

In terms of the redshift $z$, the density evolution equation of the baryon
density in the electric type superconducting dark energy models is given by
\begin{equation}
\frac{d\rho (z)}{dz}=\frac{3}{1+z}\frac{p(z)+\rho (z)\left[1-\left(\alpha
^2/4\lambda \right)\rho (z)\right]}{1-\left(\alpha ^2/4\lambda \right)\rho
(z)}.
\end{equation}

In order to characterize the dark energy, and its evolution properties, we
also introduce the dark energy equation of state parameter $w_{DE}$, defined
as
\begin{equation}
w_{DE}=\frac{p_{DE}}{\rho _{DE}}=-\frac{-\left(\alpha ^2/8\lambda\right)\rho
^2+V_0}{\left(\alpha ^2/8\lambda \right)\rho ^2+V_0}.
\end{equation}

\subsubsection{Dust Universes with electric type superconducting dark energy}

In the case of the dust Universe, with $p=0$, Eq.~(\ref{rho1}), describing
the time dynamics of the matter density in the presence of the
superconducting electric type dark energy takes the form
\begin{equation}  \label{rho2}
\dot{\rho}=-\sqrt{3}\rho \sqrt{\rho -\left( \alpha ^{2}/8\lambda \right)
\rho ^{2}+V_{0}}.
\end{equation}

By introducing a set of dimensionless variables $\left( \theta ,\tau
,h,v_{0}\right) $, defined as $\rho =\left( 8\lambda /\alpha ^{2}\right)
\theta $, $t= \alpha /\sqrt{24\lambda }\tau $, $H=\left( \sqrt{8\lambda }/%
\sqrt{3}\alpha \right) h$, $v_{0}=\left( \alpha ^{2}/8\lambda \right) V_{0}$%
, Eq.~(\ref{rho2}) becomes
\begin{equation}  \label{rho3n}
\frac{d\theta }{d\tau }=-\theta \sqrt{\theta -\theta ^{2}+v_{0}},
\end{equation}%
while the dimensionless Hubble parameter $h$ is obtained as
\begin{equation}
h=\sqrt{\theta -\theta ^{2}+v_{0}}.
\end{equation}

In the dimensionless time variable $\tau $ we have $h(\tau )=\left[3/a(\tau )%
\right](da/d\tau )$. The deceleration parameter of this model is given by
\begin{equation}
q=\frac{1}{2}\frac{\theta -4\theta ^{2}-2v_{0}}{\theta -\theta ^{2}+v_{0}},
\end{equation}
while the dark energy equation of state parameter $w_{DE}$ can be obtained
as
\begin{equation}
w_{DE}=-\frac{-\theta ^2+v_0}{\theta ^2+v_0}.
\end{equation}

The time variation of the redshift $z$ can be obtained from the equation
\begin{equation}
\frac{dz}{d\tau }=-(1+z)\frac{1}{a}\frac{da}{d\tau }=-\frac{1+z}{3}\sqrt{%
\theta -\theta ^{2}+v_{0}}.
\end{equation}

In terms of the redshift $z$ the evolution of the dust electric type
superconducting dark energy Universe is described by the simple relations
\begin{equation}
\rho (z)=\rho _{0}\left( 1+z\right) ^{3},
\end{equation}%
\begin{equation}
H(z)=\frac{1}{\sqrt{3}}\left[ \rho _{0}(1+z)^{3}-\frac{\alpha ^{2}\rho
_{0}^{2}}{8\lambda }(1+z)^{6}+V_{0}\right] ^{1/2},
\end{equation}

\begin{equation}
q(z)=\frac{1}{2}\frac{\rho _{0}(1+z)^{3}-\left( \alpha ^{2}\rho
_{0}^{2}/2\lambda \right) \left( 1+z\right) ^{6}-2V_{0}}{\rho
_{0}(1+z)^{3}-\left( \alpha ^{2}\rho _{0}^{2}/8\lambda \right) \left(
1+z\right) ^{6}+V_{0}},
\end{equation}%
where $\rho _{0}$ is the matter density of the Universe at the present time $%
z=0$.

\begin{figure}[h]
\centering
\includegraphics[width=8cm]{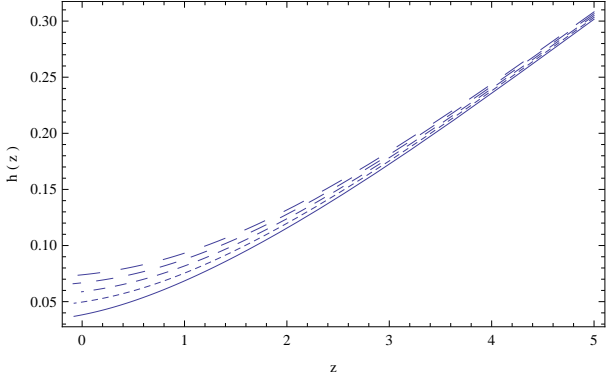}
\caption{Variation with respect to the redshift $z$ of the dimensionless
Hubble function $h$ of the electric type superconducting dark energy filled
Universe with $p=0$ for different values of $v_0$: $v_0=0.001$ (solid
curve), $v_0=0.002$ (dotted curve), $v_0=0.003$ (short dashed curve), $%
v_0=0.004$ (dashed curve), and $v_0=0.005$ (long dashed curve),
respectively. }
\label{fig1}
\end{figure}

\begin{figure}[h]
\centering
\includegraphics[width=8cm]{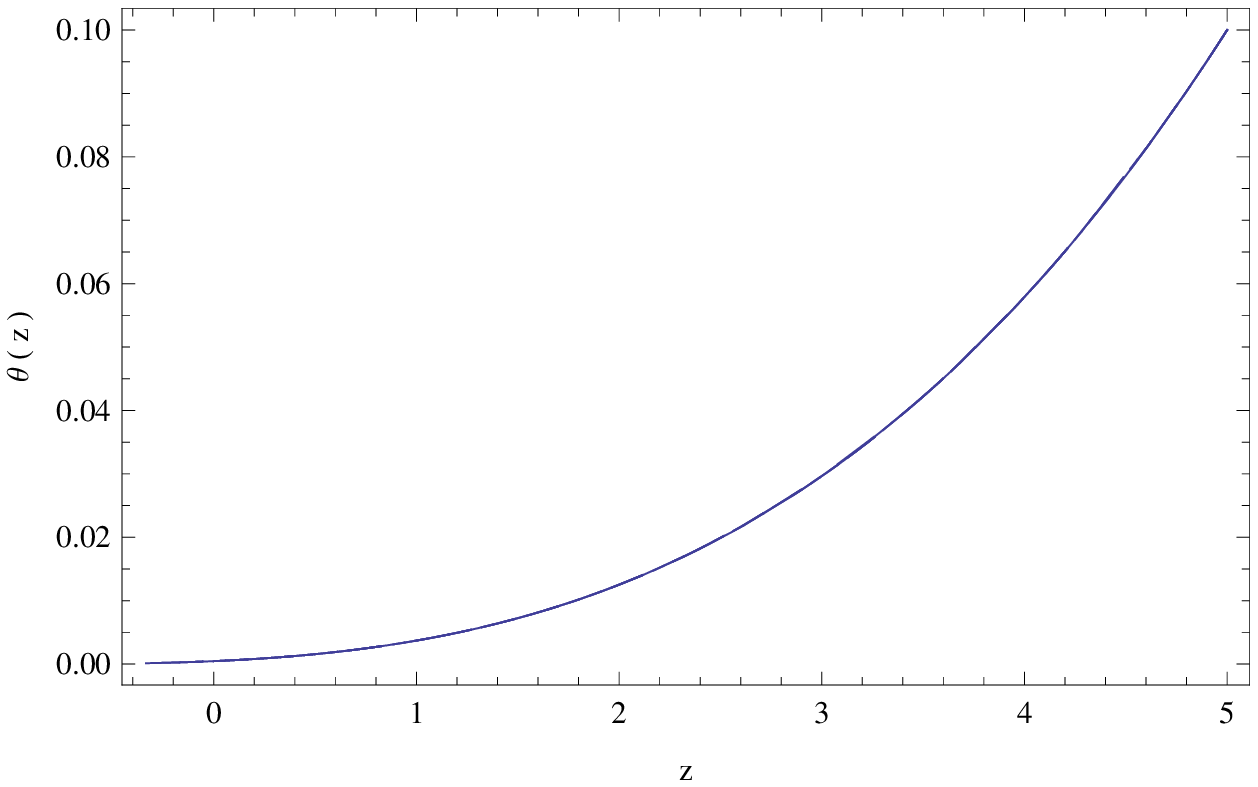}
\caption{Variation with respect to the redshift $z$ of the dimensionless
matter energy density $\theta $ of the electric type superconducting
dark energy filled Universe with $p=0$ for different values of $v_0$: $%
v_0=0.001$ (solid curve), $v_0=0.002$ (dotted curve), $v_0=0.003$ (short
dashed curve), $v_0=0.004$ (dashed curve), and $v_0=0.005$ (long dashed
curve), respectively. }
\label{fig2}
\end{figure}

\begin{figure}[h]
\centering
\includegraphics[width=8cm]{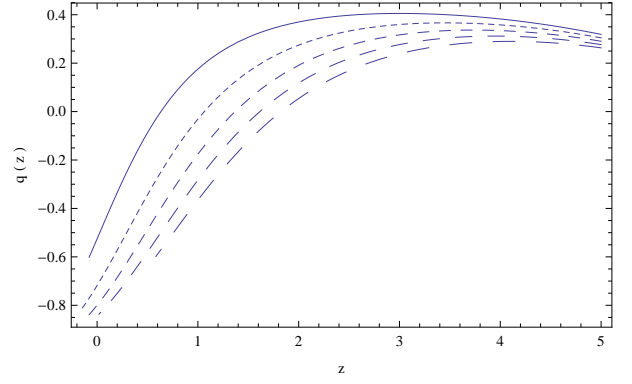}
\caption{Variation with respect to the redshift $z$ of the deceleration
parameter $q$ of the electric type superconducting dark energy filled
Universe with $p=0$ for different values of $v_0$: $v_0=0.001$ (solid
curve), $v_0=0.002$ (dotted curve), $v_0=0.003$ (short dashed curve), $%
v_0=0.004$ (dashed curve), and $v_0=0.005$ (long dashed curve),
respectively. }
\label{fig3}
\end{figure}

\begin{figure}[h]
\centering
\includegraphics[width=8cm]{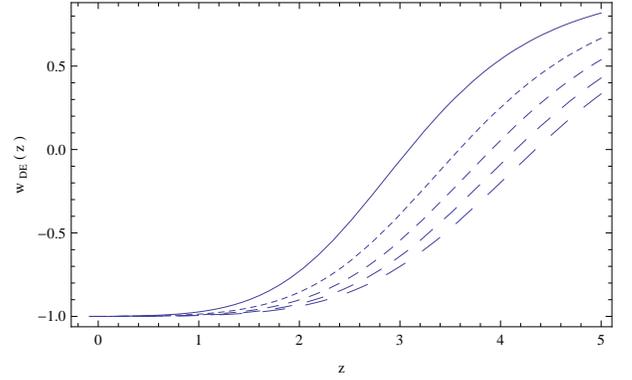}
\caption{Variation with respect to the redshift $z$ of the parameter $w_{DE}$
of the dark energy equation of state of the electric type superconducting
dark energy filled Universe with $p=0$ for different values of $v_0$: $%
v_0=0.001$ (solid curve), $v_0=0.002$ (dotted curve), $v_0=0.003$ (short
dashed curve), $v_0=0.004$ (dashed curve), and $v_0=0.005$ (long dashed
curve), respectively. }
\label{fig4}
\end{figure}

The variations with respect to the redshift $z $ of the Hubble function $h$
of the Universe, of the matter energy density $\theta $, of the deceleration
parameter $q$, and of the parameter of the dark energy equation of state are
represented, for different values of $v_0$, in Figs.~\ref{fig1}-\ref{fig4}.
The initial values used to numerically integrate the cosmological evolution
equations are $\theta (0)=0.1$, $z(0)=5$, and $a(0)=1/6$, respectively.

As one can see from Fig.~\ref{fig1}, the Hubble function $h$ of the Universe
is a monotonically increasing function of the redshift (monotonically time
decreasing function). In the early stages of evolution, at around $%
z\approx 5$, $h$ is basically independent on the values of $v_0$. The matter
density of the Universe, presented in Fig.~\ref{fig2}, is a monotonically
increasing function of $z$,  tending in the small $z$ limit to zero, $%
\lim_{z\rightarrow 0}\theta (z)=0$. Its evolution is basically independent
of the range if the numerical values of $v_0$. The redshift variation of the
deceleration parameter $q$, depicted in Fig.~\ref{fig3}, shows that in the
present model the Universe starts from a decelerating phase at around $%
z\approx 5$, with $q$ having values of the order of $q\approx 0.2-0.3$. This
initial value increases in the early stages of the cosmological evolution,
showing a decelerating expansion. At $z \approx 1-2$, the Universe starts to
accelerate, with the decelerating parameter slightly decreasing and taking
negative vales $q<0$. The values of the deceleration parameter gradually
decrease with decreasing $z$, and the Universe enters into an accelerating
stage, ending its evolution in a de Sitter stage, with $q\approx -1$ at $%
z\approx 0$. The parameter $w_{DE}$ of the equation of state of the dark
energy, presented in Fig.~\ref{fig4}, starts with positive values, and, with
decreasing $z$, it takes negative values. In the small redshift limit it
tends to the value $w_{DE}=-1$.

The present day numerical values of the Hubble function $H_{0}$ and of the
deceleration parameter $q_{0}$ can be obtained as
\begin{equation}
H_{0}=\frac{\left[ \rho _{0}-\left( \alpha ^{2}\rho _{0}^{2}/8\lambda
\right) +V_{0}\right] ^{1/2}}{\sqrt{3}},
\end{equation}
and
\begin{equation}
q_{0}=\frac{1}{2}\frac{\left[ \rho _{0}-\left( \alpha ^{2}\rho
_{0}^{2}/2\lambda \right) -2V_{0}\right] }{\left[ \rho _{0}-\left( \alpha
^{2}\rho _{0}^{2}/8\lambda \right) +V_{0}\right] },
\end{equation}
respectively. Therefore the free parameters $\alpha $ and $\lambda $ of the
superconducting electric type dark energy model can be obtained from
astronomical observations.

Eq.~(\ref{rho3n}) can also be solved exactly, and thus we obtain the density
as a function of time in an exact analytical form as given by
\begin{eqnarray}
\theta (\tau ) &=&4\theta _{0}v_{0}e^{\tau \sqrt{v_{0}}}\Bigg[ 2\sqrt{%
v_{0}\left( -\theta _{0}^{2}+ \theta _{0}+v_{0}\right) e^{2\tau _{0}\sqrt{%
v_{0}}}}+  \nonumber \\
&&(\theta _{0}+2v_{0})e^{\tau _{0}\sqrt{v_{0}}}\Bigg] \Bigg\{ \left[ \theta
_{0}^{2}+8v_{0}^{2}-4(\theta _{0}-2)\theta _{0}v_{0}\right]\times  \nonumber
\\
&&e^{2\tau _{0}\sqrt{v_{0}}}-4\theta _{0}e^{\tau \sqrt{v_{0}}}\sqrt{%
v_{0}\left( -\theta _{0}^{2}+\theta _{0}+v_{0}\right) e^{2\tau _{0}\sqrt{%
v_{0}}}}+  \nonumber \\
&&\theta _{0}^{2}(4v_{0}+1)e^{2\tau \sqrt{v_{0}}}+4(\theta
_{0}+2v_{0})e^{\tau _{0}\sqrt{v_{0}}}\times  \nonumber \\
&&\sqrt{v_{0}\left( -\theta _{0}^{2}+\theta _{0}+v_{0}\right) e^{2\tau _{0}%
\sqrt{v_{0}}}}-  \nonumber \\
&&2\theta _{0}(\theta _{0}+2v_{0})e^{\sqrt{v_{0}}(\tau +\tau _{0})}\Bigg\} %
^{-1},
\end{eqnarray}
where we have used the initial condition $\theta \left(\tau _0\right)=\theta
_0$.

\subsubsection{The radiation fluid Universe with electric type
superconducting dark energy}

For a high density radiation fluid Universe, with matter equation of state
satisfying the condition $p=\rho /3$, in the presence of electric type
superconducting dark energy the basic evolution equation of the
dimensionless matter density $\theta $ is given by
\begin{equation}
\frac{d\theta }{d\tau }=-\sqrt{\theta -\theta ^{2}+v_{0}}\frac{\theta
/3+\theta (1-2\theta )}{(1-2\theta )}.
\end{equation}
The deceleration parameter of the radiation Universe can be obtained as
\begin{equation}
q=\frac{2\theta -4\theta ^2-2v_0}{2\left(\theta -\theta ^2+v_0\right)}.
\end{equation}

\begin{figure}[h]
\centering
\includegraphics[width=8cm]{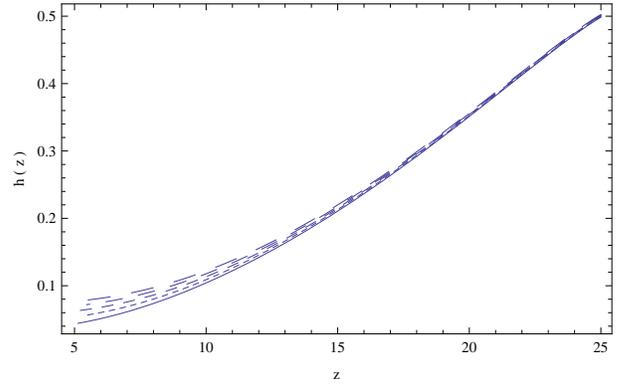}
\caption{Variation with respect to the redshift $z$ of the Hubble function $h
$ of the radiation fluid Universe with $p=\rho/3$ in the presence of
electric type superconducting dark energy for different values of $v_0$: $%
v_0=0.001$ (solid curve), $v_0=0.002$ (dotted curve), $v_0=0.003$ (short
dashed curve), $v_0=0.004$ (dashed curve), and $v_0=0.005$ (long dashed
curve), respectively. }
\label{fig5}
\end{figure}

\begin{figure}[h]
\centering
\includegraphics[width=8cm]{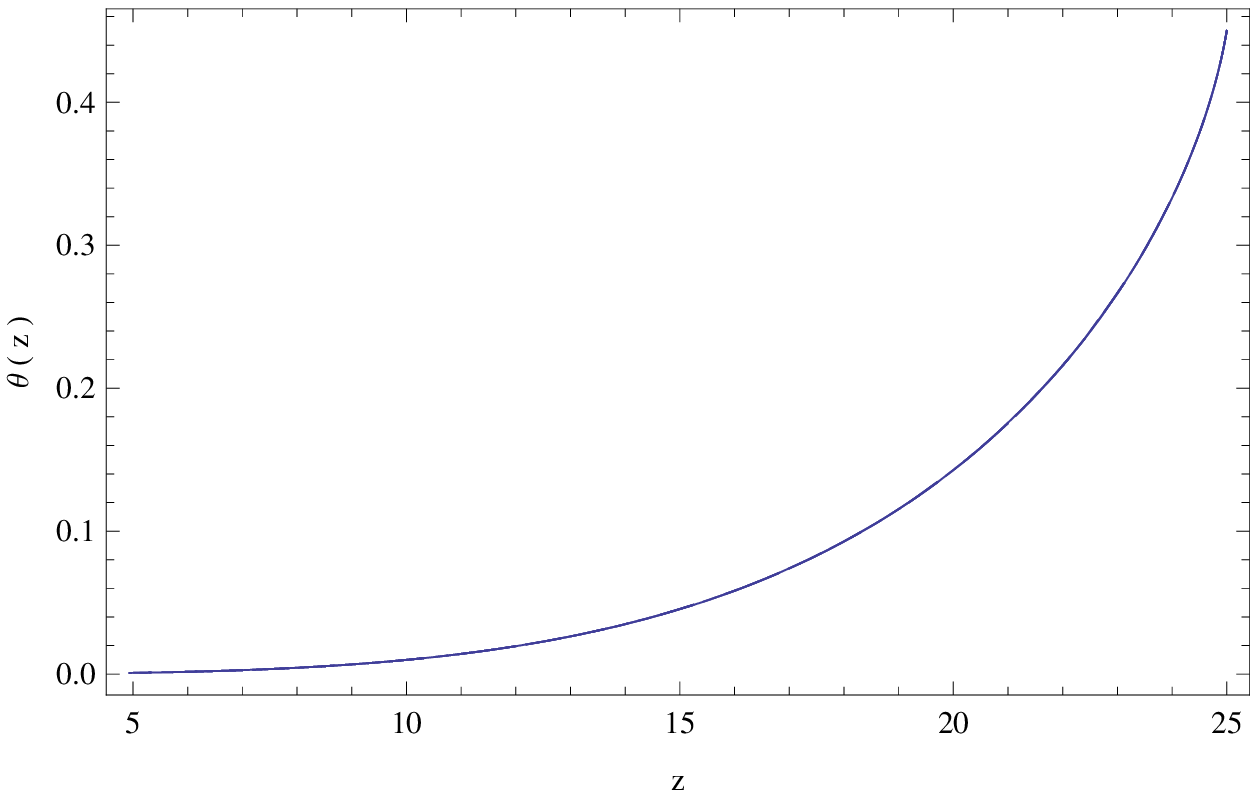}
\caption{Variation with respect to the redshift $z$ of the matter energy
density $\theta $ of the radiation fluid Universe with $p=%
\rho/3$ in the presence of electric type superconducting dark energy for
different values of $v_0$: $v_0=0.001$ (solid curve), $v_0=0.002$ (dotted
curve), $v_0=0.003$ (short dashed curve), $v_0=0.004$ (dashed curve), and $%
v_0=0.005$ (long dashed curve), respectively. }
\label{fig6}
\end{figure}

\begin{figure}[h]
\centering
\includegraphics[width=8cm]{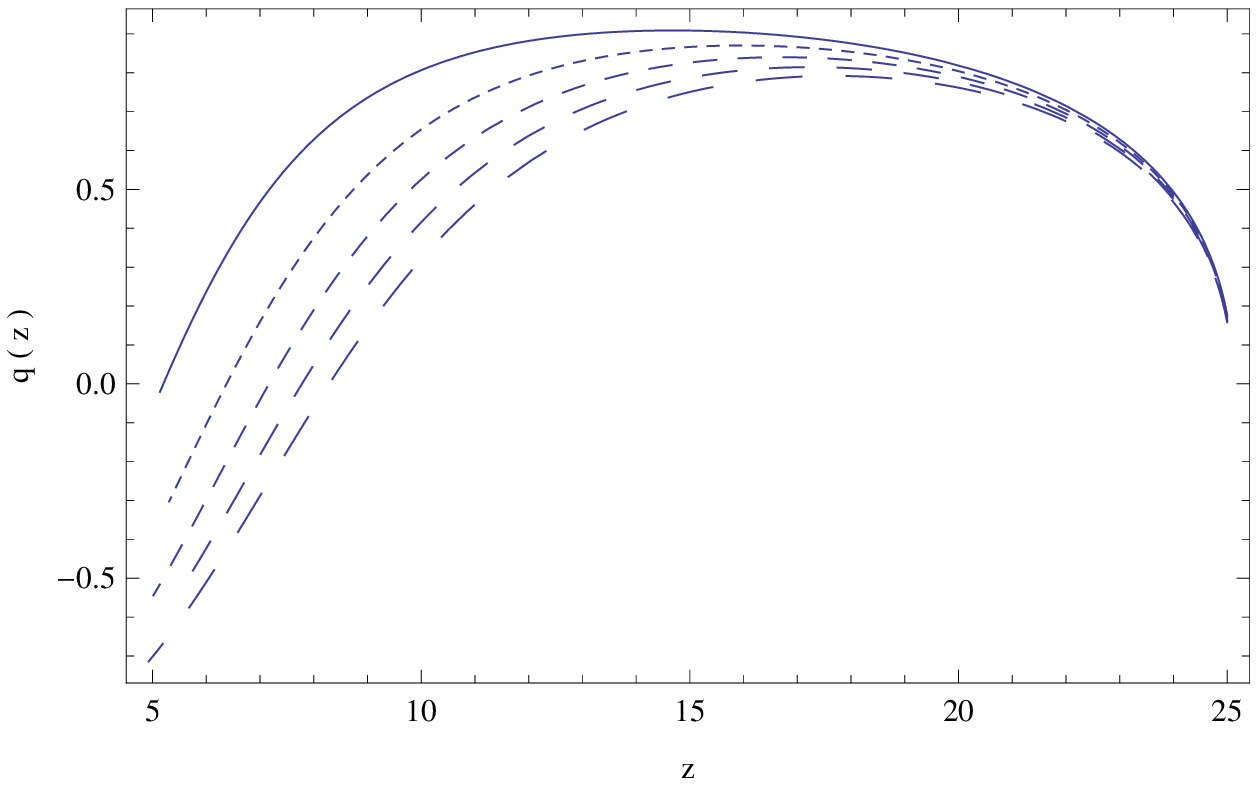}
\caption{Variation with respect to the redshift $z$ of the deceleration
parameter $q$ of the radiation fluid Universe with $p=\rho/3$ in the
presence of electric type superconducting dark energy for different values
of $v_0$: $v_0=0.001$ (solid curve), $v_0=0.002$ (dotted curve), $v_0=0.003$
(short dashed curve), $v_0=0.004$ (dashed curve), and $v_0=0.005$ (long
dashed curve), respectively. }
\label{fig7}
\end{figure}

\begin{figure}[h]
\centering
\includegraphics[width=8cm]{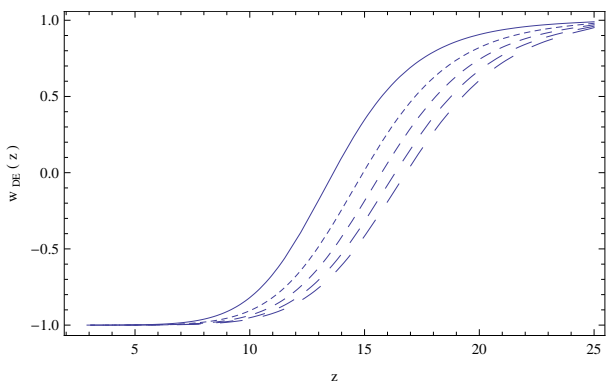}
\caption{Variation with respect to the redshift $z$ of the parameter of the
dark energy equation of state $w_{DE}$ of the radiation fluid Universe with $%
p=\rho/3$ in the presence of electric type superconducting dark
energy for different values of $v_0$: $v_0=0.001$ (solid curve), $v_0=0.002$
(dotted curve), $v_0=0.003$ (short dashed curve), $v_0=0.004$ (dashed
curve), and $v_0=0.005$ (long dashed curve), respectively. }
\label{fig8}
\end{figure}

The variations with respect to the redshift $z$ of the Hubble function, of
the energy density, of the deceleration parameter and of the parameter of
the dark energy equation of state of the radiation fluid Universe in the
presence of the electric type superconducting dark energy are presented in
Figs.~\ref{fig5}-\ref{fig8}. The initial conditions used to numerically
integrate the cosmological evolution equation are $\theta (0)=0.45$ and $%
z(0)=25$.

We assume that the Universe was radiation dominated in the redshift range $%
5\leq z\leq 25$. The dimensionless Hubble function $h$ of the high density
Universe, presented in Fig.~\ref{fig5}, is a monotonically increasing
function of the redshift (monotonically decreasing in time), while the
energy density, depicted in Fig.~\ref{fig6}, increases monotonically with
the redshift $z$ during the cosmological evolution. The deceleration
parameter, shown in Fig.~\ref{fig7}, has positive values for the redshift
interval $9\leq z \leq 25$, indicating a decelerating expansion. For large
values of $v_0$ the deceleration parameter can reach the zero value at
redshifts as high as $z\approx 9$, $\left.\lim_{z \rightarrow
9}q\right|_{v_0=0.005}\approx 0$. The time variation of the cosmological
parameters $h$ and $\theta $ is practically independent of the adopted small
values of the parameter $v_0$. The parameter $w_{DE}$ of the dark energy
equation of state is represented in Fig.~\ref{fig8}. For large redshift
values $15\leq z\leq 25$, $w_{DE}$ is positive, while for $z\approx 5$ it
approaches the value $w_{DE}=-1$, showing that in the present model the
Universe becomes dark energy dominated at around $z\approx 5$.

\subsubsection{The unified picture of the evolution of the Universe in the electric type
superconducting dark energy model}

\begin{figure}[h]
\centering
\includegraphics[width=8cm]{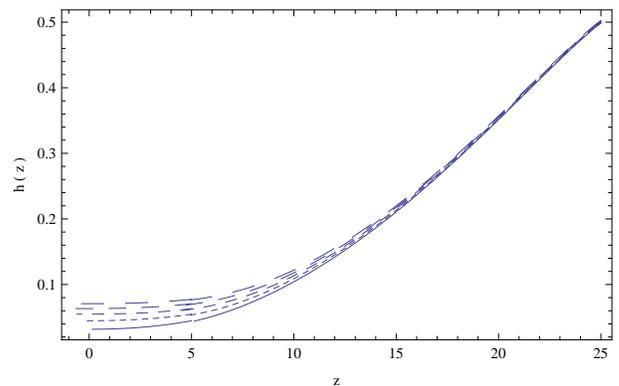}
\caption{Variation with respect to the redshift $z\in[0,25]$ of the Hubble function $h
$ of the Universe filled by
electric type superconducting dark energy, for different values of $v_0$: $%
v_0=0.001$ (solid curve), $v_0=0.002$ (dotted curve), $v_0=0.003$ (short
dashed curve), $v_0=0.004$ (dashed curve), and $v_0=0.005$ (long dashed
curve), respectively. }
\label{fign1}
\end{figure}

\begin{figure}[h]
\centering
\includegraphics[width=8cm]{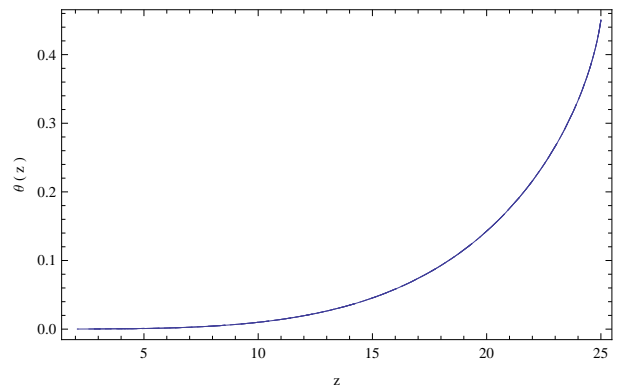}
\caption{Variation with respect to the redshift $z\in [0,25]$ of the matter energy
density $\theta $ of the Universe filled with electric type superconducting dark energy, for
different values of $v_0$: $v_0=0.001$ (solid curve), $v_0=0.002$ (dotted
curve), $v_0=0.003$ (short dashed curve), $v_0=0.004$ (dashed curve), and $%
v_0=0.005$ (long dashed curve), respectively. }
\label{fign2}
\end{figure}

\begin{figure}[h]
\centering
\includegraphics[width=8cm]{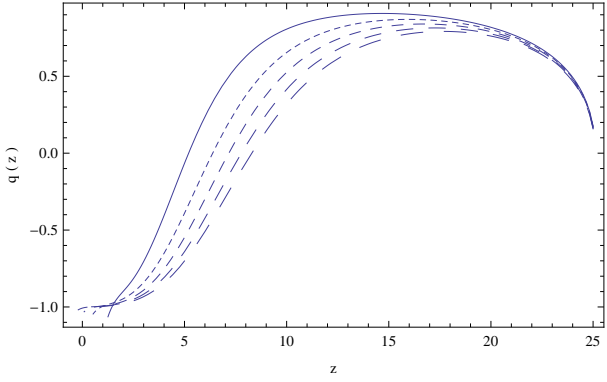}
\caption{Variation with respect to the redshift $z\in [0,25]$ of the deceleration
parameter $q$ of the Universe filled with electric type superconducting dark energy, for different values
of $v_0$: $v_0=0.001$ (solid curve), $v_0=0.002$ (dotted curve), $v_0=0.003$
(short dashed curve), $v_0=0.004$ (dashed curve), and $v_0=0.005$ (long
dashed curve), respectively. }
\label{fign3}
\end{figure}

\begin{figure}[h]
\centering
\includegraphics[width=8cm]{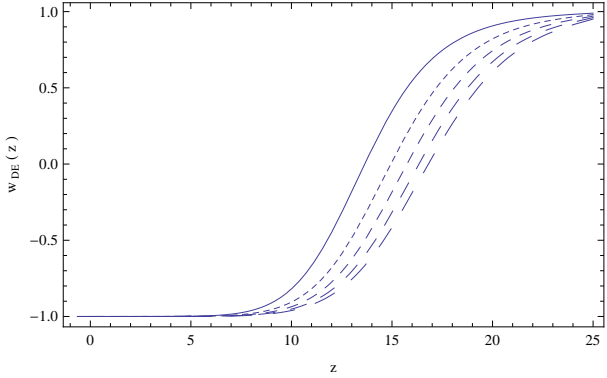}
\caption{Variation with respect to the redshift $z\in [0,25]$ of the parameter of the
dark energy equation of state $w_{DE}$ for an Universe filled with electric type superconducting dark
energy for different values of $v_0$: $v_0=0.001$ (solid curve), $v_0=0.002$
(dotted curve), $v_0=0.003$ (short dashed curve), $v_0=0.004$ (dashed
curve), and $v_0=0.005$ (long dashed curve), respectively. }
\label{fign4}
\end{figure}

Finally, to conclude the investigation of the electric type superconducting dark energy model, we present a unified picture of the evolution of the Universe for the redshift range $z\in[0,25]$. The variations of the Hubble function, matter energy density, deceleration parameter, and the parameter of the dark energy equation of state are plotted in Figs.~\ref{fign1}-\ref{fign4}.

To study the evolution of the electric type superconducting dark energy we adopt for the redshift $z$ the range from 0 to 25. We assume that in the range $z\in [5,25]$ the matter content of the Universe can be (at least approximately) described by a radiation type equation of state $p=\rho /3$. At $z=5$ the Universe enters in the matter dominated era, with $p\approx 0$. In this simplified model the transition from the radiation dominated era to the matter dominated phase is smooth, with all physical and thermodynamical quantities continue at the transition point. Therefore the Hubble function and the matter density, represented in Figs.~\ref{fign1} and \ref{fign2}, are monotonically increasing functions of the redshift for the entire period. The deceleration parameter, shown in Fig.~\ref{fign3}, has a complex behavior. The Universe starts at $z=25$ with a deceleration parameter having a value of $q\approx 0.10$, and its early evolution is strongly decelerating, with $q$ reaching the value $q\approx 0.9$ at $z\approx 12$, for $v_0=0.001$. For larger values of $v_0$ the expansion of the Universe is faster, with $q$ reaching the value $q=0.6$ at $z\approx 14$. After $q$ has reached its maximum, it starts to decrease with decreasing $z$, and, depending on the numerical value of $v_0$,  reaches the value $q=0$ for $z\approx 5-7$. Then the deceleration parameter enters the negative range, with the Universe starting to accelerate at a higher rate, and reaching the value $q\approx -1$ (the de Sitter phase) at around $z=0$. The parameter of the dark energy equation of state $w_{DE}$, represented in Fig.~\ref{fign4}, is slowly decreasing from its maximum value 1 in the redshift range $z\approx 17-25$. For $z<15$, $w_{DE}$ decreases rapidly with decreasing $z$, and, depending on the numerical value of $v_0$,  reaches the limiting value $w_{DE}\approx -1$ at $z\approx 7.5-10$.

\subsubsection{Electric type superconducting dark energy Universe with
conserved electric field and matter current}

Finally, we investigate the case in which the electric potential $A^0$ and
the matter current satisfy the conservation equations given by Eqs.~(\ref%
{cond1}). The time variation of $A^0$ can be immediately obtained from the
Lorentz gauge equation imposed on $A^{\mu}$, which gives
\begin{equation}
\nabla _{\mu}A^{\mu}=\frac{1}{\sqrt{-g}}\frac{\partial }{\partial x^{\mu}}%
\left(\sqrt{-g}A^{\mu}\right)=\frac{1}{a^3}\frac{d}{dt}\left(a^3A^0\right)=0,
\end{equation}
and
\begin{equation}
A^0(t)=\frac{C_0}{a^3},
\end{equation}
respectively, where $C_0$ is an arbitrary integration constant. The
continuity equation of the matter hydrodynamic flow, $\nabla
_{\mu}\left(\rho u^{\mu}\right)=0$ gives a similar dependence for the matter
density $\rho $,
\begin{equation}  \label{20n}
\rho (t)=\frac{\rho _0}{a^3},
\end{equation}
where $\rho _0$ is an arbitrary integration constant. The evolution of the
scalar field $\phi $ is decoupled from the electric and matter component,
and in the presence of a constant potential, $V\left(A^2,\phi\right)=\mathrm{%
constant}$, follows a similar law as the electric and the matter fields,
\begin{equation}
\phi (t)=\frac{\phi _0}{a^3},
\end{equation}
where $\phi _0$ is an arbitrary integration constant. In the large time
limit, all these fields tend to zero, and the Universe enters in an
exponential, de Sitter type expansionary phase. However, the presence of the
electric type superconducting dark energy modifies the cosmological dynamics
of the Universe before it enters in the de Sitter stage.

\subsection{Magnetic dark energy models}

As a second superconducting vector type dark energy model we consider the
case in which dark energy has a magnetic type structure, with its vector
potential given by $A_{\mu}=\left(0, A_1(t),A_2(t),A_3(t)\right)$.
In order to have an isotropic expansion the components of the
superconducting magnetic vector potential must satisfy the condition $%
A_1(t)=A_2(t)=A_3(t)=A(t)$. Hence for this choice of $A_{\mu}$ we obtain $%
F_{i0}=-\dot{A}_i$, $F^{i0}=\dot{A}_i/a^2$, $i=1,2,3$, $\left(1/16\pi%
\right)F_{\alpha \beta}F^{\alpha \beta}=-\left(3/8\pi\right)\dot{A}^2/a^2$, $%
\left(1/4\pi\right)F^0_{\alpha}F^{\alpha }_0=-\left(3/4\pi\right)\dot{A}%
^2/a^2$, and $(1/4\pi )F^i_{\alpha}F^{\alpha }_i=-\left(1/4\pi\right)\dot{A}%
/a^2$ (no summation upon the index $i$). Therefore the gravitational field
equations describing the isotropic and homogeneous Universe in the presence
of superconducting dark energy take the form
\begin{equation}
3H^2=\left(1-\frac{\alpha }{2}\dot{\phi}\right)\rho +\frac{\lambda }{2}\dot{%
\phi}^2+\frac{3}{8\pi}\frac{\dot{A}^2}{a^2}+\frac{3\lambda }{2}\frac{A^2}{a^2%
}+V_0,
\end{equation}
\begin{equation}
2\dot{H}+3H^2=-p+\frac{\alpha }{2}\rho \dot{\phi}-\frac{\lambda }{2}\dot{\phi%
}^2-\frac{1}{8\pi }\frac{\dot{A}^2}{a^2}+\frac{\lambda }{2}\frac{A^2}{a^2}%
+V_0,
\end{equation}
\begin{equation}  \label{49}
\lambda \dot{\phi}-\frac{\alpha }{2}\rho =0,
\end{equation}
\begin{equation}
\frac{1}{a}\frac{d}{dt}\left(a\dot{A}\right)=-4\pi \lambda A,
\end{equation}
where for simplicity we have adopted a constant value $V_0$ for the
self-interaction potential of the scalar and vector fields, $%
V\left(A^2,\phi\right)=V_0=\mathrm{constant}$. The energy conservation
equation takes the form
\begin{eqnarray}  \label{eqcn}
&&\frac{d}{dt}\left( a^{3}\rho \right) +p\frac{d}{dt}\left( a^{3}\right) =%
\frac{\alpha }{2}a^{3}\dot{\phi}\dot{\rho }+\frac{15}{8\pi }\dot{a}\dot{A}%
^{2}+\frac{7}{3}\lambda \dot{a}A^{2}-  \nonumber \\
&&3\left( \frac{1}{2}\lambda \dot{\phi}^{2}+V_{0}\right) a^{2}\dot{a}.
\end{eqnarray}

With the use of Eq.~(\ref{49}) we can substitute the derivative of the
scalar field in terms of the matter density $\rho $. Therefore the system of
gravitational field equations describing the superconducting magnetic type
cosmological dark energy model takes the form
\begin{equation}
3\frac{\dot{a}^{2}}{a^{2}}=\rho -\frac{\alpha ^{2}}{8\lambda }\rho ^{2}+%
\frac{3}{8\pi }\frac{\dot{A}^{2}}{a^{2}}+\frac{3\lambda }{2}\frac{A^{2}}{%
a^{2}}+V_{0},  \label{52}
\end{equation}%
\begin{equation}
2\frac{\ddot{a}}{a}+\frac{\dot{a}^{2}}{a^{2}}=-p+\frac{\alpha ^{2}}{8\lambda
}\rho ^{2}-\frac{1}{8\pi }\frac{\dot{A}^{2}}{a^{2}}+\frac{\lambda }{2}\frac{%
A^{2}}{a^{2}}+V_{0},  \label{53}
\end{equation}%
\begin{equation}
\ddot{A}+\frac{\dot{a}}{a}\dot{A}+4\pi \lambda A=0.  \label{54n}
\end{equation}%
From Eqs.~(\ref{52}) and (\ref{53}) we obtain
\begin{equation}
2\dot{H}=-(\rho +p)+\frac{\alpha ^{2}}{4\lambda }\rho ^{2}-\frac{1}{2\pi }%
\frac{\dot{A}^{2}}{a^{2}}-\lambda \frac{A^{2}}{a^{2}}.
\end{equation}%
The deceleration parameter of the magnetic type superconducting dark energy
model can be represented as
\begin{widetext}
\begin{equation}
q=\frac{\rho +3p-\left( \alpha ^{2}/2\lambda \right) \rho ^{2}+\left( 3/4\pi
\right) \left( \dot{A}^{2}/a^{2}\right)
-2V_{0}}{2\left[ \rho -\left( \left( \alpha ^{2}/8\lambda \right) \rho
^{2}\right) +\left( 3/8\pi \right) \left( \dot{A}^{2}/a^{2}\right) +\left(
3\lambda /2\right) \left( A^{2}/a^{2}\right) +V_{0}\right] }.
\end{equation}
\end{widetext}while the parameter of the equation of state of the dark
energy is given by
\bea
&&\hspace{-0.5cm}w_{DE}=\nonumber\\
&&\hspace{-0.5cm}-\frac{-\left( \alpha ^{2}/8\lambda \right) \rho ^{2}+\left( 3/8\pi
\right) \left( \dot{A}^{2}/a^{2}\right) +\left( 3\lambda /2\right) \left(
A^{2}/a^{2}\right) +V_{0}}{\left( \alpha ^{2}/8\lambda \right) \rho
^{2}-\left( 1/8\pi \right) \left( \dot{A}^{2}/a^{2}\right) +\left( \lambda
/2\right) \left( A^{2}/a^{2}\right) +V_{0}}.\nonumber\\
\eea

In the case of a dust Universe, with $p=0$, from the conservation equation
Eq.~(\ref{eqcn}) we obtain for the time derivative of the energy density $%
\rho$ the equation
\begin{equation}
\dot{\rho }=\frac{\frac{15}{8\pi }\frac{\dot {A}^{2}}{a^{2}}+\frac{7}{3}%
\lambda \frac{A^{2}}{a^{2}}-\frac{3\alpha ^{2}}{8\lambda }\rho ^{2}+3\rho
-3V_{0}}{\left( 1-\frac{\alpha ^{2}}{4\lambda }a^{3}\rho \right) }H.
\end{equation}

\subsubsection{Dust Universes with magnetic type superconducting dark energy}

For the case of dust matter, with negligible thermodynamic pressure, we can
take $p=0$ in the gravitational field equations. Then by introducing a set
of dimensionless variables $\left(\theta ,\Sigma, \tau, h,v_0\right)$
defined as $\left\{\rho =\left(8\lambda /\alpha ^2\right)\theta, A=\sqrt{%
8\pi /3}\Sigma,t=\left(\alpha/\sqrt{8\lambda }\right)\tau,\right.$ $%
\left.H=\left(\sqrt{8\lambda}/\alpha \right)h,V_0=\left(8\lambda /\alpha
^2\right)V_0 \right\}$, and by denoting $\sigma =\pi \alpha ^2/2$, the
system of Eqs.~(\ref{52})-(\ref{54n}) can be written in a dimensionless form
as
\begin{equation}  \label{g1}
3h^2=\theta -\theta ^2+\frac{1}{a^2}\left(\frac{d\Sigma }{d\tau}%
\right)^2+\sigma \frac{\Sigma ^2}{a^2}+v_0,
\end{equation}
\begin{equation}  \label{58}
2\frac{dh}{d\tau}+3h^2=\theta ^2-\frac{1}{3}\frac{1}{a^2}\left(\frac{d\Sigma
}{d\tau }\right)^2+\frac{\sigma }{3}\frac{\Sigma ^2}{a^2}+v_0,
\end{equation}
\begin{equation}  \label{g2}
2\frac{dh}{d\tau}=\theta (2\theta -1)-\frac{4}{3}\frac{1}{a^2}\left(\frac{%
d\Sigma }{d\tau}\right)^2-\frac{2}{3}\sigma \frac{\Sigma ^2}{a^2},
\end{equation}
\begin{equation}  \label{59}
\frac{d^2\Sigma }{d\tau ^2}+h\frac{d\Sigma }{d\tau }+\sigma \Sigma =0,
\end{equation}
\begin{equation}  \label{59a}
\frac{dz}{d\tau}=-(1+z)\frac{1}{\sqrt{3}}\sqrt{\theta -\theta ^2+\frac{1}{a^2%
}\left(\frac{d\Sigma }{d\tau}\right)^2+\sigma \frac{\Sigma ^2}{a^2}+v_0}.
\end{equation}

The deceleration parameter and the parameter of the equation of state of the dark energy of the Universe filled with magnetic type
superconducting dark energy are obtained as
\begin{equation}
q=\frac{\theta -4\theta ^2+2\Sigma ^{\prime 2}/a^2-2v_0}{2\left(\theta
-\theta ^2+\Sigma ^{\prime 2}/a^2+\sigma \Sigma ^2/a^2+v_0\right)},
\end{equation}
\be
w_{DE}=-\frac{-\theta ^2+\Sigma ^{\prime 2}/a^2
+\sigma \Sigma ^2/a^2+v_0}{\theta ^2-\Sigma ^{\prime 2}/3a^2+\sigma \Sigma ^2/3a^2+v_0},
\ee
where a prime denotes the derivative with respect to the dimensionless time $%
\tau $. By taking the derivative with respect to $\tau $ of Eq.~(\ref{g1}),
and with the use of Eqs.~(\ref{g2}) and (\ref{59}) we obtain for the time
variation of the matter density the equation
\begin{equation}  \label{60}
\theta =\frac{\theta _0}{a^3},
\end{equation}
where $\theta _0$ is an arbitrary constant of integration.

\begin{figure}[h]
\centering
\includegraphics[width=8cm]{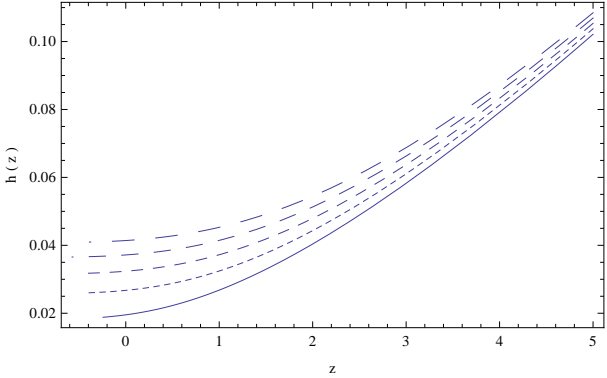}
\caption{The Hubble function of the Universe in the presence of magnetic
type superconducting dark energy as a function of redshift for $%
\sigma =0.0001$, and for different values of $v_0$: $v_0=0.001$ (solid
curve), $v_0=0.002$ (dotted curve), $v_0=0.003$ (short dashed curve), $v_0=0.004
$ (dashed curve), and $v_0=0.005$ (long dashed curve), respectively. }
\label{fig9}
\end{figure}

The variations with respect of the redshift $z\in [0,3]$ of the Hubble
function, of the matter energy density $\theta $, and of the deceleration
parameter $q$ of the Universe filled with magnetic type superconducting dark
energy, obtained by numerically integrating Eqs.~(\ref{58}), (\ref{59}), (%
\ref{59a}), and (\ref{60}), are presented, for a fixed value of $\sigma
=0.0001$, and for different values of $v_0$, in Figs.~\ref{fig7}-\ref{fig9}.
The initial conditions use to integrate the system of cosmological evolution
equations are $\theta (0)=\theta _0=0.25 $, $a(0)=1$, $\Sigma (0)=0.01$, $%
z(0)=5$, and $\Sigma ^{\prime }(0)=0.01$, respectively.

\begin{figure}[h]
\centering
\includegraphics[width=8cm]{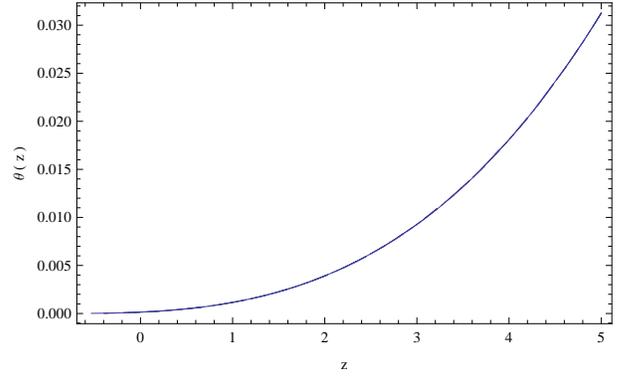}
\caption{ Matter energy density of the dust Universe in the presence of
magnetic type superconducting dark energy as a function of the redshift for $%
\sigma =0.0001$, and for different values of $v_0$: $v_0=0.001$
(solid curve), $v_0=0.002$ (dotted curve), $v_0=0.003$ (short dashed curve), $%
v_0=0.004$ (dashed curve), and $v_0=0.005$ (long dashed curve), respectively. }
\label{fig10}
\end{figure}

\begin{figure}[h]
\centering
\includegraphics[width=8cm]{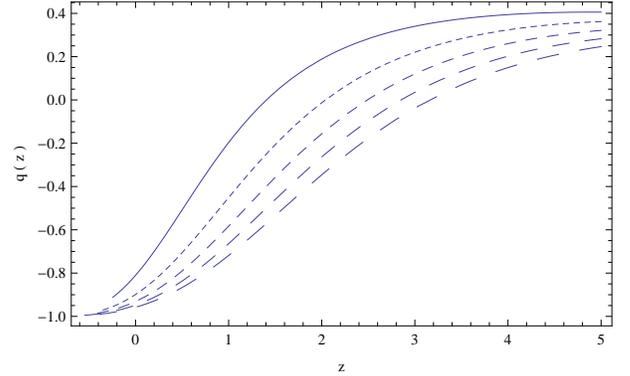}
\caption{Redshift evolution of the deceleration parameter of the dust
Universe in the presence of magnetic type superconducting dark energy for $%
\sigma =0.0001$, and for different values of $v_0$: $v_0=0.001$
(solid curve), $v_0=0.002$ (dotted curve), $v_0=0.003$ (short dashed curve), $%
v_0=0.004$ (dashed curve), and $v_0=0.005$ (long dashed curve), respectively. }
\label{fig11}
\end{figure}

\begin{figure}[h]
\centering
\includegraphics[width=8cm]{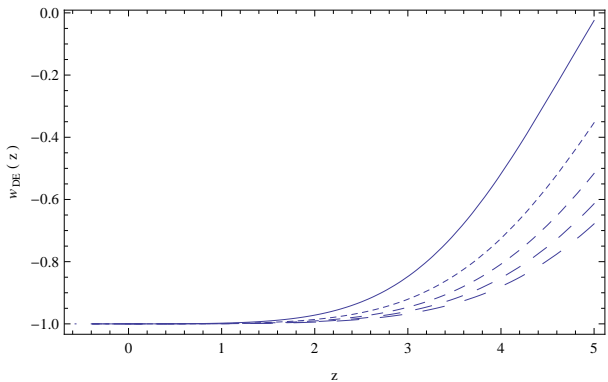}
\caption{Redshift evolution of the  parameter of the dark energy equation of state of the dust
Universe in the presence of magnetic type superconducting dark energy for $%
\sigma =0.0001$, and for different values of $v_0$: $v_0=0.001$
(solid curve), $v_0=0.002$ (dotted curve), $v_0=0.003$ (short dashed curve), $%
v_0=0.004$ (dashed curve), and $v_0=0.005$ (long dashed curve), respectively. }
\label{fig12}
\end{figure}

As one can see from Fig.~\ref{fig9}, the Hubble function of the Universe filled with magnetic type superconducting dark energy is a monotonically
increasing function of $z$ (time decreasing function), indicating an
expansionary evolution. The matter energy density $\theta$, represented in
Fig.~\ref{fig10}, monotonically increases with the redshift, and tends to
zero in the limit of small $z$. Its dynamics is basically independent on the
adopted numerical values of the parameters $\sigma $ and $v_0$. The dust magnetic Universe
starts from a decelerating state at $z=5$, with positive values of the
deceleration parameter $q>0$, shown in Fig.~\ref{fig11}. The cosmological evolution is generally decelerating for $2\leq z\leq 5$, with $q$ reaching the value zero at $z\approx 2$. Then the Universe begins to accelerate, with $q<0$,  and in the large time (small $z$) limit we have $\lim
_{z \rightarrow 0}q(z)=-1$. Thus, in the final stages of evolution of the
Universe fillet with magnetic type superconducting dark energy the
cosmological expansion is of de Sitter type, with the dark energy driving the Universe's expansion. The parameter of the dark energy equation of state, represented in Fig.~\ref{fig12}, is smaller than zero in the entire redshift range $0\leq z\leq 5$, and it tends to -1 in the limit of small redshifts.

\section{Thermodynamic interpretation of the superconducting dark energy
models}

\label{therm}

In the present Section we analyze the physical interpretation of the
superconducting dark energy model by adopting the point of view of the
thermodynamics of the matter creation irreversible processes \cite{Pri0}-%
\cite{Lima}. As we have already seen, the energy conservation equation of
the superconducting dark energy models, Eq.~(\ref{em2}), contain, as
compared to the standard adiabatic conservation equation, an extra term,
which can be interpreted thermodynamically as a matter creation rate.
According to irreversible thermodynamics, matter creation represents an
entropy source, generating an entropy flux, and thus modifying the
temperature evolution of the considered gravitational system. On the other
hand, due to our choice of the geometry of the Universe, all the
non-diagonal components of the total energy--momentum tensor of the
superconducting dark energy model are equal to zero, so that $T_{\mu
\nu}^{(total)}=0$, $\mu \neq \nu $. In particular, from the point of
view of the thermodynamics of the irreversible processes and open systems, this condition
implies the impossibility of heat transfer in the
Friedmann--Robertson--Walker models of superconducting dark energy, since
the condition $T_{0i}^{(total)}\equiv 0$, $i=1,2,3$ must always hold.

\subsection{Matter creation rates and the creation pressure}

To analyze the thermodynamical implications of the superconducting dark
energy models at the cosmological scale we start with an open system containing $N$
particles in a volume $V=a^{3}$, and characterized by an energy density $%
\rho $ and a thermodynamic pressure $p$. For such a system the second law of
thermodynamics, in its most general form, is given by \cite{Pri}
\begin{equation}
\frac{d}{dt}\left( \rho a^{3}\right) +p\frac{d}{dt}a^{3}=\frac{dQ}{dt}+\frac{%
\rho +p}{n}\frac{d}{dt}\left( na^{3}\right) ,  \label{21}
\end{equation}%
where $dQ$ is the heat received by the system during time $dt$, and $n=N/V$
is the particle number density, respectively. Due to our choice of the
geometry of the Universe, in a homogeneous and isotropic system filled with superconducting dark energy only
adiabatic transformations, defined by the condition $dQ=0$, are possible.
Therefore in the following we ignore proper heat transfer processes in the
superconductor type cosmological system. However, under the assumption of adiabatic
transformations, Eq.~(\ref{21}), representing the general formulation of the second
law of thermodynamics, contains the term $[(\rho +p)/n]d\left( na^{3}\right)
/dt$, which explicitly takes into account the variation of the number of
particles in a given volume. Hence, in the general thermodynamic approach of open
systems, even for the case of adiabatic transformations with $dQ=0$, there
is a "heat" (internal energy), received/lost by the system, which is entirely due to
the change in the particle number $n$. From the cosmological perspective of
the superconducting dark energy models, the change in the particle number is
due to the transfer of energy from dark energy to matter. Thus in this class of cosmological models matter
creation acts as a source of internal energy, as well as of entropy. For adiabatic
transformations $dQ/dt=0$, Eq.~(\ref{21}) can be written in an equivalent
form as
\begin{equation}
\dot{\rho}+3(\rho +p)H=\frac{\rho +p}{n}\left( \dot{n}+3Hn\right) .
\label{cons0}
\end{equation}

Therefore, from the point of view of the thermodynamics of open systems,
Eq.~(\ref{em2}), giving the energy conservation equation in the
superconducting dark energy models, can be interpreted as describing
particle creation in an homogeneous and isotropic geometry, with the time
variation of the particle number obtained from the equation
\begin{equation}
\dot{n}+3nH=\Gamma n,  \label{22}
\end{equation}%
where the particle creation rate $\Gamma $ is defined as
\begin{eqnarray}  \label{33}
\Gamma &=&\frac{1}{\rho +p}\Bigg\{ -\frac{\alpha }{2}u^{\nu }\nabla _{\mu }%
\left[ j^{\mu }\left( A_{\nu }-\nabla _{\nu }\phi \right) \right] +
\nonumber \\
&&\frac{\alpha }{2}\frac{d}{ds}\left[ j^{\beta }\left( A_{\beta }-\nabla
_{\beta }\phi \right) \right] -\partial _{\phi }V\left( A^{2},\phi \right)
u^{\nu }A_{\nu }-  \nonumber \\
&&2\partial _{A^{2}}V\left( A^{2},\phi \right) u^{\nu }A^{\alpha }\nabla
_{\alpha }A_{\nu }\Bigg\} .
\end{eqnarray}

Therefore the energy conservation equation in the superconducting dark
energy model can be written in the alternative form
\begin{equation}
\dot{\rho}+3(\rho +p)H=(\rho +p)\Gamma .  \label{41}
\end{equation}

As shown initially in \cite{Pri}, for adiabatic transformations Eq.~(\ref{21}%
), describing irreversible particle creation in an open thermodynamic
systems, can be formulated as an effective energy conservation equation of
the form
\begin{equation}
\frac{d}{dt}\left( \rho a^{3}\right) +\left( p+p_{c}\right) \frac{d}{dt}%
a^{3}=0,
\end{equation}%
which can be written in an equivalent form as,
\begin{equation}
\dot{\rho}+3\left( \rho +p+p_{c}\right) H=0,  \label{comp}
\end{equation}%
where we have introduced the term $p_{c}$, called the creation pressure, and
which is defined as \cite{Pri}
\begin{eqnarray}
p_{c} &=&-\frac{\rho +p}{n}\frac{d\left( na^{3}\right) }{da^{3}}=-\frac{\rho
+p}{3nH}\left( \dot{n}+3nH\right) =  \nonumber \\
&&-\frac{\rho +p}{3}\frac{\Gamma }{H}.
\end{eqnarray}%
Therefore in the superconducting dark energy model the creation pressure can
be obtained as
\begin{eqnarray}  \label{pc}
p_{c} &=&-\frac{1}{3H}\Bigg\{-\frac{\alpha }{2}u^{\nu }\nabla _{\mu }\left[
j^{\mu }\left( A_{\nu }-\nabla _{\nu }\phi \right) \right] +  \nonumber \\
&&\frac{\alpha }{2}\frac{d}{ds}\left[ j^{\beta }\left( A_{\beta }-\nabla
_{\beta }\phi \right) \right] -\partial _{\phi }V\left( A^{2},\phi \right)
u^{\nu }A_{\nu }-  \nonumber \\
&&2\partial _{A^{2}}V\left( A^{2},\phi \right) u^{\nu }A^{\alpha }\nabla
_{\alpha }A_{\nu }\Bigg\}.
\end{eqnarray}

\subsubsection{Particle creation rates and creation pressure in the electric type superconducting dark energy model}

As an example of the thermodynamic description of the superconducting dark
energy models we consider the electric type superconducting dark energy
case, for which the energy conservation equation is given by Eq.~(\ref{cex}%
), can be formulated as
\begin{equation}
\dot{\rho}+3(\rho +p)H=\frac{\alpha ^{2}}{4\lambda }\rho \left( \dot{\rho}%
+3H\rho \right) .  \label{c1}
\end{equation}

From Eq.~(\ref{c1}) it follows that for $p=0$ the matter energy is conserved,
$\dot{\rho}+3(\rho +p)H=0$, and there is no particle creation from the
superconducting dark energy. However, for $p\neq 0$, matter and energy
transfer processes take place in the presence of the superconducting
electric type dark energy, with the particle creation rate $\Gamma $ given
by
\begin{equation}  \label{86}
\Gamma =\frac{\alpha ^{2}}{4\lambda }\frac{\rho }{\rho +p}\left( \dot{\rho}%
+3H\rho \right) .
\end{equation}

The creation pressure for this model can be obtained as
\begin{equation}
p_{c}=-\frac{\alpha ^{2}}{12\lambda }\frac{\rho }{H}\left( \dot{\rho}+3H\rho
\right) .
\end{equation}

\subsection{Entropy and temperature evolution}

In order to formulate the second law of thermodynamics for open systems, and to apply it to the superconducting dark energy model,   we
must decompose the entropy change in the cosmological fluid into two components: the entropy flow term
$d_eS$, and the entropy creation term $d_iS$. Hence the total entropy $S$ of
an open thermodynamic system can be written as \cite{Pri0,Pri}
\begin{equation}
dS = d_eS + d_iS,
\end{equation}
where we assume that $d_iS > 0$. Both the entropy flow and the entropy production rate in the superconducting dark energy model
can be evaluated by starting from the total differential of the entropy,
given by \cite{Pri},
\begin{equation}
T d\left(\bar{s}a^3\right)=d\left(\rho a^3\right)+pda^3-\mu
d\left(na^3\right),
\end{equation}
where $T$ is the temperature of the open thermodynamic system with superconducting particle creation, $\bar{s}%
=S/a^3 $ is the entropy per unit volume, and $\mu $ is the chemical
potential, defined in the usual way as
\begin{equation}
\mu n=(\rho +p)-T\bar{s}.
\end{equation}

For closed systems and adiabatic transformations $dS=0$ and $d_iS=0$.
However, in the presence of matter creation there is a non-zero contribution
to the entropy. For homogeneous systems the entropy flow term $d_eS$
cancels, so that $d_eS = 0$. But matter creation from superconducting dark energy acts as a source for
entropy production, with the corresponding entropy time variation obtained
as \cite{Pri}
\begin{eqnarray}  \label{25}
T\frac{d_iS}{dt}&=&T\frac{dS}{dt}=\frac{\rho +p}{n}\frac{d}{dt}%
\left(na^3\right)-\mu \frac{d}{dt}\left(na^3\right)=  \nonumber \\
&&T\frac{\bar{s}}{n}\frac{d}{dt}\left(na^3\right)\geq 0,
\end{eqnarray}

From Eq.~(\ref{25}) we obtain for the time variation of the entropy the
equation
\begin{equation}
\frac{dS}{dt}=\frac{S}{n}\left( \dot{n}+3Hn\right) =\Gamma S\geq 0,
\label{43}
\end{equation}%
giving for the entropy increase due to particle creation the expression
\begin{equation}
S(t)=S_{0}e^{\int_{0}^{t}{\Gamma \left( t^{\prime }\right) dt^{\prime }}},
\end{equation}%
where $S_{0}=S(0)$ is a constant. With the use of Eq.~(\ref{86}), we obtain
for the entropy production in the superconducting dark energy models the
equation
\begin{eqnarray}
\frac{1}{S}\frac{dS}{dt} &=&\frac{1}{\rho +p}\Bigg\{-\frac{\alpha }{2}u^{\nu
}\nabla _{\mu }\left[ j^{\mu }\left( A_{\nu }-\nabla _{\nu }\phi \right) %
\right] +  \nonumber  \label{entff} \\
&&\frac{\alpha }{2}\frac{d}{ds}\left[ j^{\beta }\left( A_{\beta }-\nabla
_{\beta }\phi \right) \right] -\partial _{\phi }V\left( A^{2},\phi \right)
u^{\nu }A_{\nu }-  \nonumber \\
&&2\partial _{A^{2}}V\left( A^{2},\phi \right) u^{\nu }A^{\alpha }\nabla
_{\alpha }A_{\nu }\Bigg\}\geq 0.
\end{eqnarray}

Equivalently, the above equation can be written as
\begin{equation}
\frac{1}{S}\frac{dS}{dt}=\frac{\alpha ^{2}}{4\lambda }\frac{\rho }{\rho +p}%
\left( \dot{\rho}+3H\rho \right) \geq 0.
\end{equation}

An important thermodynamic quantity, the entropy flux vector $S^{\mu }$ of the particles created from the superconducting dark energy, is
defined according to \cite{Cal}
\begin{equation}
S^{\mu }=n\sigma u^{\mu },
\end{equation}%
where $\sigma =S/N$ is the specific entropy per particle. The entropy flux
vector $S^{\mu }$ must satisfy during the entire cosmological evolution the second law of thermodynamics, which
requires that the constraint $\nabla _{\mu }S^{\mu }\geq 0$ be satisfied for all times. By taking into
account the fundamental Gibbs relation \cite{Cal},
\begin{equation}
nTd\sigma =d\rho -\frac{\rho +p}{n}dn,
\end{equation}%
and by using the definition of the chemical potential $\mu $ of the
superconducting thermodynamic system as given by
\begin{equation}
\mu =\frac{\rho +p}{n}-T\sigma ,
\end{equation}%
we obtain
\begin{eqnarray}
\nabla _{\mu }S^{\mu } &=&\left( \dot{n}+3nH\right) \sigma +nu^{\mu }\nabla
_{\mu }\sigma =  \nonumber  \label{48} \\
&&\frac{1}{T}\left( \dot{n}+3Hn\right) \left( \frac{\rho +p}{n}-\mu \right) ,
\end{eqnarray}%
where we have taken into account the important relation
\begin{equation}
nT\dot{\sigma}=\dot{\rho}-\frac{\rho +p}{n}\dot{n}=0,
\end{equation}%
which follows immediately from Eq.~(\ref{cons0}). With the use of Eq.~(\ref%
{86}) we obtain for the entropy production rate due to the particle creation
processes in the superconducting dark energy model the expression
\begin{eqnarray}  \label{efn}
\nabla _{\mu }S^{\mu } &=&\frac{n}{\left( \rho +p\right) T}\Bigg\{-\frac{%
\alpha }{2}u^{\nu }\nabla _{\mu }\left[ j^{\mu }\left( A_{\nu }-\nabla _{\nu
}\phi \right) \right] +  \nonumber \\
&&\frac{\alpha }{2}\frac{d}{ds}\left[ j^{\beta }\left( A_{\beta }-\nabla
_{\beta }\phi \right) \right] -\partial _{\phi }V\left( A^{2},\phi \right)
u^{\nu }A_{\nu }-  \nonumber \\
&&2\partial _{A^{2}}V\left( A^{2},\phi \right) u^{\nu }A^{\alpha }\nabla
_{\alpha }A_{\nu }\Bigg\}\left( \frac{\rho +p}{n}-\mu \right) .  \nonumber \\
\end{eqnarray}

The entropy production rate via superconducting particle creation processes given by Eq.~(%
\ref{efn}) can be written in a simpler form as
\begin{equation}
\nabla _{\mu }S^{\mu }=\frac{\alpha ^{2}}{4\lambda }\frac{1}{T}n\left( \frac{%
\rho +p}{n}-\mu \right) \frac{\rho }{\rho +p}\left( \dot{\rho}+3H\rho
\right).
\end{equation}

In the general case the thermodynamic state of a perfect comoving fluid is
described by only two essential thermodynamic variables, the particle number
density $n$, and the temperatures $T$, respectively. Hence the energy density
$\rho $ and the thermodynamic pressure $p$ can be obtained, in terms of the particle number $n$
and temperature $T$, by using the standard  form of the equilibrium equations of state of the matter created by the superconducting dark energy,
\begin{equation}
\rho =\rho (n,T),p=p(n,T).  \label{51}
\end{equation}%
Therefore the energy conservation equation Eq.~(\ref{41}) can be written in the expanded
form
\begin{equation}
\frac{\partial \rho }{\partial n}\dot{n}+\frac{\partial \rho }{\partial T}%
\dot{T}+3(\rho +p)H=(\rho +p)\Gamma .
\end{equation}%
With the use of the general thermodynamic relation \cite{Cal}
\begin{equation}
\frac{\partial \rho }{\partial n}=\frac{\rho +p}{n}-\frac{T}{n}\frac{%
\partial p}{\partial T},
\end{equation}%
we obtain for the temperature evolution of the newly created particles in
the superconducting dark energy model the expression
\begin{equation}
\frac{\dot{T}}{T}=c_{s}^{2}\frac{\dot{n}}{n}=c_{s}^{2}\left( \Gamma
-3H\right) .  \label{54}
\end{equation}%
where the speed of sound $c_{s}$ is defined as $c_{s}^{2}=\partial
p/\partial \rho $. If the matter newly created from the superconducting dark energy  satisfies a barotropic
equation of state $p=\left( \gamma -1\right) \rho $, $1\leq \gamma \leq 2$,
the matter temperature evolution can be obtained as
\begin{equation}
T=T_{0}n^{\gamma -1}.
\end{equation}

\section{Discussions and final remarks}

\label{sect3}

In the present paper we have considered an electromagnetic type dark energy
model, in which the electromagnetic gauge invariance is spontaneously
broken. The action for such a system must be invariant under gauge
transformations, $A_{\mu}(x)\rightarrow A_{\mu }(x)+\partial _{\mu}\Lambda
(x)$, $\psi _n(x)\rightarrow \exp \left(iq_n\Lambda (x)/\hbar \right)\psi
_n(x)$, where $q_n$ are the charges destroyed by the field $\psi _n$ \cite{W}%
. These phase changes lead to the formation of an ordered state. By writing
all charged fields as a function of a scalar field $\phi (x)$, when the
matter fields are integrated out we obtain a Lagrangian that is a gauge
invariant functional of the fields $A^{\mu}$ and $\phi $. Such a physical
model can explain easily all the observed properties of superconductors \cite%
{W}-\cite{F}. Tentatively, we also propose it as a dark energy model with a
broken electromagnetic gauge invariance.

From a physical point of view the superconducting dark energy model is a
two-field model, leading to a scalar-vector-tensor cosmological theory. It
can also be viewed as a unified scalar - vector field dark energy model, in
which the scalar field $\phi $ and the vector field $A_{\mu}$ appear in the
gauge invariant combination $A_{\mu}-\nabla _{\mu}\phi$. Moreover, similarly
the standard electrodynamic case, we have assumed the possible existence of
a generalized coupling between the matter current $j^{\mu}$ and the gauge
invariant combination of the potentials $A_{\mu}-\nabla _{\mu}\phi$. We have
investigated the cosmological implications of this model, by restricting our
analysis to the case of a homogeneous and isotropic geometry. We have
considered two distinct classes of models, whose main properties are
determined by the form of the electromagnetic potential $A_{\mu}$. The first
model corresponds to an electric type choice for the dark energy potential,
with $A_0$ the only non-zero component. For this case the general solution
of the gravitational field equations was obtained numerically for the dust
and the radiation filled Universe, respectively. In both cases we have
assumed that the self-interaction potential of the electromagnetic type dark
energy is a constant. In both cases in the long time limit the Universe ends
in a decelerating phase. A similar result is obtained for the magnetic type
dark energy model, in which the electromagnetic potential is restricted to
the form $A_{\mu}=\left(0,A(t),A(t),A(t)\right)$. For this case we have
investigated, by numerically solving the gravitational field equations, the
zero thermodynamic pressure (dust) cosmological model, in the presence of a
constant self-interaction potential. Similarly to the electric type dark
energy model, the magnetic superconducting dark energy model drives the
Universe, in the long time limit, in an accelerated, de Sitter type,
expansionary phase.

Due to the coupling between the dark energy potentials and the matter
current, the matter energy-momentum tensor is not conserve in the present
approach. We have interpreted, in the framework of the thermodynamics of
open systems and irreversible processes \cite{Pri0}-\cite{Lima}, the
non-conservation of the matter energy-momentum tensor as describing particle
creation, and energy transfer from the superconducting dark energy to
ordinary matter. We have explicitly obtained the particle creation rates, as
well as the effective creation pressure generated by the irreversible
transformation of the field energy into matter. The entropy production rate
and the overall entropy evolution of the Universe was also obtained, with
the total entropy being given by the exponential of the time integral of the
particle production rate $\Gamma $.

The possible observational study in a cosmological context of the
irreversible matter--creation processes in the homogeneous and isotropic
flat Friedmann--Robertson--Walker geometry in the superconducting dark
energy models may represent one of the possibilities of considering the
viability of this dark energy model. However, in order to confirm the
validity of the superconducting dark energy model developed in the present
paper, it is necessary to carefully consider a much wider range of
cosmological and astrophysical tests for this type of models. In particular,
an essential test of the superconducting dark energy model would be the
investigation of the classical macroscopic predictions of the model in
large-scale structure formation with linear perturbations, and with the
consideration of the Newtonian limit for small scales. Supernovae
observations fitting, and the study of the effects of the matter creation on
the Cosmic Microwave Background anisotropies could lead to other important
tests and parameter constraints of the model. Essentially the
superconducting dark energy model introduced in the present paper is a
simple toy model, whose main goal is to stimulate the study of alternative
or more general electromagnetic type dark energy models.

The superconducting dark energy model introduced in the present paper leads
to the possibility that matter creation, associated with matter current -
dark energy electromagnetic potential coupling may also happen in the
present - day universe, as initially considered by Dirac \cite{22}. The
existence of some forms of coupling between matter and dark energy are not
in contradiction with the cosmological observations or with some
astrophysical data \cite{Boe}. However, firm observational evidence for
particle creation on a cosmological scale is still missing. Hopefully, a key
ingredient of the present model, the functional form of the self-interaction
potential $V\left( A^2,\phi \right)$, which essentially determines the
cosmological dynamics in the superconducting dark energy model, will be
provided by fundamental particle physics (or perhaps even condensed matter)
models, thus permitting an in depth comparison of the predictions of the
model with high precision observational cosmological and astrophysical data.

\section*{Acknowledgments}

S.-D. L. gratefully acknowledges financial support for this project from the
Fundamental Research Fund of China for the Central Universities. T. H. would
like to thank the Department of Physics of the Sun-Yat Sen University in
Guangzhou, P. R. China, for the kind hospitality offered during the
preparation of this work. %

\appendix

\section{The divergence of the energy-momentum tensor}

\label{app1}

In order to obtain the divergence of the matter energy-momentum tensor in
the superconducting dark energy model we compute first the divergence of the
electromagnetic type term,
\begin{equation}
T_{\nu }^{\left( em\right) \mu }=\frac{1}{4\pi }\left( -F_{\nu \alpha
}F^{\mu \alpha }+\frac{1}{4}F_{\alpha \beta }F^{\alpha \beta }\delta _{\nu
}^{\mu }\right) .
\end{equation}

Hence for the divergence of the electromagnetic component we find first
\begin{equation}
\nabla _{\mu }T_{\nu }^{\left( em\right) \mu }=\frac{1}{4\pi }\left( \frac{1%
}{2}F^{\alpha \beta }\nabla _{\nu }F_{\alpha \beta }-\nabla _{\mu }F_{\nu
\alpha }F^{\mu \alpha }-F_{\nu \alpha }\nabla _{\mu }F^{\mu \alpha }\right) .
\end{equation}

By taking into account that $\left( 1/4\pi \right) \nabla _{\mu }F^{\mu
\alpha }=J^{\alpha }$, and $\nabla _{\nu }F_{\alpha \beta }=-\nabla _{\alpha
}F_{\beta \nu }-\nabla _{\beta }F_{\nu \alpha }$, it follows that
\begin{eqnarray}
\nabla _{\mu }T_{\nu }^{\left( em\right) \mu }&=&\frac{1}{4\pi }\Bigg( -%
\frac{1}{2}F^{\alpha \beta }\nabla _{\alpha }F_{\beta \nu }-\frac{1}{2}%
F^{\alpha \beta }\nabla _{\beta }F_{\nu \alpha }-  \nonumber \\
&&F^{\mu \alpha }\nabla _{\mu }F_{\nu \alpha }\Bigg) -F_{\nu \alpha
}J^{\alpha }.
\end{eqnarray}

The terms in the bracket vanish, and therefore we find
\begin{eqnarray}
\nabla _{\mu }T_{\nu }^{\left( em\right) \mu }&=&F_{\nu \alpha }\Bigg[ %
\lambda g^{\alpha \beta }\left( A_{\beta }-\nabla _{\beta }\phi \right) +%
\frac{\alpha }{2}j^{\alpha }-  \nonumber \\
&&2\partial _{A^{2}}V\left( A^{2},\phi \right) A^{\alpha }\Bigg] .
\end{eqnarray}

Then for the divergence of the matter energy-momentum tensor we obtain
\begin{eqnarray}
&&\hspace{-0.4cm}\nabla _{\mu }T_{\nu }^{\mu }-F_{\nu \alpha }J^{\alpha
}+\lambda \left( \nabla _{\mu }A^{\mu }-\nabla _{\mu }\nabla ^{\mu }\phi
\right) \left( A_{\nu }-\nabla _{\nu }\phi \right) +  \nonumber \\
&&\hspace{-0.4cm}\lambda \left( A^{\mu }-\nabla ^{\mu }\phi \right) F_{\mu
\nu }+\alpha \nabla _{\mu }j^{\mu }\left( A_{\nu }-\nabla _{\nu }\phi
\right) +  \nonumber \\
&&\hspace{-0.4cm}\frac{\alpha }{2}j^{\mu }\left( \nabla _{\mu }A_{\nu
}-\nabla _{\mu }\nabla _{\nu }\phi \right) + \frac{\alpha }{2}j^{\mu }F_{\mu
\nu }-\frac{\alpha }{2}\nabla _{\nu }j^{\beta }\left( A_{\beta }-\nabla
_{\beta }\phi \right) +  \nonumber \\
&&\hspace{-0.4cm}\partial _{\phi }V\left( A^{2},\phi \right) \nabla _{\nu
}\phi +2A^{\alpha }\partial _{\phi }V\left( A^{2},\phi \right) \nabla _{\nu
}A_{\alpha }=0,
\end{eqnarray}
where we have used the identities
\begin{eqnarray}
&&\hspace{-0.5cm}\lambda \left( A^{\mu }-\nabla ^{\mu }\phi \right) \left(
\nabla _{\mu }A_{\nu }-\nabla _{\mu }\nabla _{\nu }\phi \right) -\frac{%
\lambda }{2}\left( \nabla _{\nu }A^{\alpha }-\nabla _{\nu }\nabla ^{\alpha
}\phi \right)\times  \nonumber \\
&&\hspace{-0.5cm} \left(A_{\alpha }-\nabla _{\alpha }\phi \right) -\frac{%
\lambda }{2}\left( A^{\alpha }-\nabla ^{\alpha }\phi \right) \left( \nabla
_{\nu }A_{\alpha }-\nabla _{\nu }\nabla _{\alpha }\phi \right) =  \nonumber
\\
&&\hspace{-0.5cm}\lambda F_{\mu \nu }\left( A^{\mu }-\nabla ^{\mu }\phi
\right) ,
\end{eqnarray}
and
\begin{eqnarray}
&&\alpha j^{\mu }\left( \nabla _{\mu }A_{\nu }-\nabla _{\mu }\nabla _{\nu
}\phi \right) -\frac{\alpha }{2}j^{\beta }\left( \nabla _{\nu }A_{\beta
}-\nabla _{\nu }\nabla _{\beta }\phi \right) =  \nonumber \\
&&\frac{\alpha }{2}j^{\mu }\left( \nabla _{\mu }A_{\nu }-\nabla _{\mu
}\nabla _{\nu }\phi \right) +\frac{\alpha }{2}j^{\mu }F_{\mu \nu },
\end{eqnarray}
respectively.

With the use of the evolution equation of the scalar field we find the
relation,
\begin{eqnarray}
&&\hspace{-0.6cm}\lambda \left( \nabla _{\mu }A^{\mu }-\nabla _{\mu }\nabla
^{\mu }\phi \right) \left( A_{\nu }-\nabla _{\nu }\phi \right) =  \nonumber
\\
&&\hspace{-0.6cm}\partial _{\phi }V\left( A^{2},\phi \right) \left( A_{\nu
}-\nabla _{\nu }\phi \right) -\frac{\alpha }{2}\nabla _{\mu }j^{\mu }\left(
A_{\nu }-\nabla _{\nu }\phi \right) .  \label{Ap1}
\end{eqnarray}
By substituting the above relation and the expression of the divergence of
the electromagnetic part of the energy-momentum tensor in Eq.~(\ref{Ap1}),
we finally obtain
\begin{eqnarray}
&&\nabla _{\mu }T_{\nu }^{\mu }+\frac{\alpha}{2} \nabla _{\mu }\left[j^{\mu
}\left( A_{\nu }-\nabla _{\nu }\phi \right)\right] -\frac{\alpha }{2}\nabla
_{\nu }j^{\beta }\left( A_{\beta }-\nabla _{\beta }\phi \right) +  \nonumber
\\
&&\partial _{\phi }V\left( A^{2},\phi \right) A_{\nu }+ 2\partial
_{A^{2}}V\left( A^{2},\phi \right) A^{\alpha }\nabla _{\alpha }A_{\nu }=0.
\end{eqnarray}

\end{document}